\documentclass[
aps,
physrev,
twocolumn,
10pt,
]{revtex4-2}

\usepackage{graphicx}
\usepackage{amsmath}
\usepackage{natbib}
\usepackage{array}
\usepackage{amssymb}
\usepackage{color}
\usepackage{hyperref}

\usepackage{tabularray}
\UseTblrLibrary{booktabs, siunitx}

\newcommand{\Fig}[1]{Figure~\ref{#1}}
\newcommand{\Figs}[2]{Figures~\ref{#1} and \ref{#2}}

\newcommand{\Eq}[1]{Equation~(\ref{#1})}
\newcommand{\Sec}[1]{Section~\ref{#1}}

\newcommand{\Tab}[1]{Table~\ref{#1}}

\graphicspath{{./figs/}{./png/}}

\newcommand{\Msun}{\mathrm{M}_{\odot}}
\newcommand{\ee}{\mathrm{e}}
\newcommand{\aye}{\mathrm{i}}
\newcommand{\diff}{\,\mathrm{d}}

\usepackage{enumitem}  %--for itemsep in itemize.

\usepackage{acronym}

\usepackage{siunitx}
\begin{document}

\title{Auto-encoder model for faster generation of effective one-body gravitational waveform approximations}

\author{Suyog Garg}
\email{gargsuyog@g.ecc.u-tokyo.ac.jp}
\affiliation{Department of Physics, The University of Tokyo, Bunkyo-ku, Tokyo 113-0033 JAPAN}
\affiliation{Research Center for Early Universe, The University of Tokyo, Bunkyo-ku, Tokyo 113-0033 JAPAN}
% [0000-0002-2309-9731]

\author{Feng-Li Lin}
\email{fengli.lin@gmail.com}
\affiliation{Department of Physics, National Taiwan Normal University, Taipei 11677 TAIWAN}
\affiliation{Center of Astronomy and Gravitation, National Taiwan Normal University, Taipei 11677 TAIWAN}

\author{Kipp Cannon}
\affiliation{Department of Physics, The University of Tokyo, Bunkyo-ku, Tokyo 113-0033 JAPAN}
\affiliation{Research Center for Early Universe, The University of Tokyo, Bunkyo-ku, Tokyo 113-0033 JAPAN}
% \email{}


\date{\today}

\begin{abstract}
% -- background
% -- problem / motivation
Upgrades to current gravitational wave detectors for the next observation run and upcoming third-generation observatories, like the Einstein telescope, are expected to have enormous improvements in detection  sensitivities and compact object merger event rates. Estimation of source parameters for a wider parameter space that these detectable signals will lie in, will be a computational challenge.
Thus, it is imperative to have methods to speed-up the likelihood calculations with theoretical waveform predictions, which can ultimately make the parameter estimation faster and aid in rapid multi-messenger follow-ups.
% -- solution / methods
In this work we study auto-encoder models for gravitational waveform generation by adopting the best-performing architecture of \citet{liao2021} to approximate aligned-spin \texttt{SEOBNRv4} inspiral-merger-ringdown waveforms.
Our parameter space consists of four parameters, [$m_1$, $m_2$, $\chi_1(z)$, $\chi_2(z)$]. The masses are uniformly sampled in $[5,75]\,\Msun$ with a mass ratio limit at $10\,\Msun$, while the spins are uniform in $[-0.99,0.99]$. 
% -- results
Our model is able to generate $10^3$ waveforms in $\sim\qty{1e-1}{\second}$ at an average speed of about \qty{50}{\micro\second}  per waveform on a GPU. This is about 4 orders of magnitude faster than the native \texttt{SEOBNRv4} implementation, and 2--3 orders of magnitude faster than existing non-machine-learning accelerated  waveform variants. The median mismatch for the generated waveforms in the test dataset is $\sim10^{-2}$, with better performance in a restricted parameter space of $\chi_{\rm eff}\in[-0.80,0.80]$.
The latent sampling error of our model can be quantified at a median mismatch standard deviation of $4\times10^{-3}$.
% -- impact
Although the accuracy of our model does not enable full production-use yet, the model could be useful wherever high-volume of approximate theoretical waveforms are required, for instance, for rapid sky localization.
\end{abstract}

% \keywords{Gravitational-Wave -- Machine-Learning -- Low-Latency -- Auto-Encoder -- Waveform-Generation}

\maketitle
\acrodef{CBC}{compact binary coalescence}
\acrodefplural{CBC}[CBCs]{compact binary coalescences}

\acrodef{SNR}{signal-to-noise ratio}
\acrodefplural{SNR}[SNRs]{signal-to-noise ratios}

\acrodef{PDF}{probability density function}
\acrodefplural{PDF}[PDFs]{probability density functions}

\acrodef{GW}{gravitational wave}
\acrodefplural{GW}[GWs]{gravitational waves}

\acrodef{FAR}{false-alarm rate}

\acrodef{IFAR}{inverse false-alarm rate}

\acrodef{LR}{likelihood ratio}
\acrodefplural{LR}[LRs]{likelihood ratios}

\acrodef{MDA}{mock data analysis}
\acrodefplural{MDA}[MDAs]{mock data analyses}

\acrodef{LVK}{LIGO--Virgo--KAGRA}

\acrodef{CL}{curriculum learning}

\acrodef{BBH}{binary black hole}
\acrodefplural{BBH}[BBHs]{binary black holes}

\acrodef{BNS}{binary neutron star}
\acrodefplural{BNS}[BNSs]{binary neutron stars}

\acrodef{EM}{electromagnetic}
\acrodef{PE}{parameter estimation}
\acrodef{KS}{Kolmogorov-Sminov}

\acrodef{p-p}{probability-probability}

\acrodef{RESCEU}{the Research Center for the Early Universe}

\acrodef{CE}{Cosmic Explorer}

%%%%%%%%%%%%%%%%%%%%%%%%%%%%%%%%%%%%%%%%%%%%%%%%%%%%%%%%%%%%%%%%%%%%%%
%%%%%%%%%%%%%%%%%%%%%%%%%%%%%%%%%%%%%%%%%%%%%%%%%%%%%%%%%%%%%%%%%%%%%%

\section{Introduction}
\label{sec:intro}

Since the first detection of \acp{GW} in 2015 \cite{abbott2016}, the \ac{LVK} collaboration has now detected signals from more than 350 compact binary merger events of varying significance in their four observation runs \cite{abbott2019,collaboration2021,collaboration2023,theligoscientificcollaboration2025a}. The process of obtaining 
scientific outputs from each such observation event involves signal detection in noisy data and then estimation of the source parameters using Bayesian inference by calculation of noise-weighted likelihoods. For a single confident event detection, typically about a million likelihood evaluations with theoretical waveforms are required to accurately estimate the source parameter posteriors \cite{balasubramanian1996,thrane2019}. It is also required that waveforms of sufficient length be used for each likelihood calculation \cite{hannam2010}. This is especially true for GW signals from sources like 
binary neutron star systems, which tend to be have much longer waveforms than those from binary black hole sources. 
In some of the past observations, to hasten the parameter inference, part of the inspiral waveform has been truncated, so that 
the posteriors are estimated using short duration data. For instance for the detection of GW170817 a lower frequency cut-off of \qtyrange{30}{45}{\hertz} was used \cite{collaboration2017a}.

In the upcoming years, third-generation \ac{GW} observatories, such as the ground-based Einstein Telescope \cite{dellamonica2025,abac2025b} and Cosmic Explorer \cite{hall2021} or the space-based LISA \cite{colpi2024,xuan2023}, are expected to observe orders of magnitude more 
compact object merger events, with figures of around $10^4$ or more events per year often cited \cite{gupta2024}.
Even in the next O5 observation run of the current detectors, the detection sensitivity is expected to increase many times their current limits \cite{gupta2024,kagra2025}.
For each of these detected events, we will need to generate the required number of accurate theoretical waveform predictions and then compute the likelihoods, so that the source parameters can be properly constrained.
With the increased number of detections by the upcoming detectors, a more extensive parameter space coverage will be necessary to account for all the different configurations of source binary systems that can give rise to these detectable signals, for example, those with eccentric inspiral orbits and precessing binary spins. Fast and accurate theoretical waveform generation and rapid parameter estimation for gravitational waves from such complex systems is notoriously difficult and computationally expensive. Thus, we face a bottleneck in parameter estimation for detections in future observation runs.

Now, to obtain accurate theoretical waveform predictions, the source binary system can be evolved from inspiral through merger to ringdown using numerical relativity simulations \cite{scheel2025}. However, solving the full set of general relativity equations is computationally expensive as the parameter space scales or more complex sources, such as those with higher mass ratios, are included. Therefore, to make waveform computation faster, the \ac{GW} community has developed various families of gravitational waveform approximations. These are: inspiral-only Post-Newtonian expansions \cite{messina2017}, phenomenological models \cite{husa2016,khan2016,dietrich2019,pratten2020,abac2024}, effective one-body approximations \cite{buonanno1999,damour2009,nagar2018,kawaguchi2018a,mihaylov2023,meent2023,albanesi2025}, reduced order quadrature methods \cite{chua2019,morras2023}, and numerical relativity surrogates models \cite{blackman2017,blackman2017a,pathak2024,field2025,nee2025a}.

There could be two ways of making parameter estimations faster by mitigating the cost of likelihood calculations with theoretical waveforms: 1) alternate methods for likelihood calculation and parameter estimation that are faster; and 2) alternate methods for coming up with rapid predictions of the theoretical waveforms.
Relative binning \cite{zackay2018,dai2018,leslie2021,krishna2023,narola2023,narola2024}, efficient sampling algorithms and related techiniques \cite{vinciguerra2017,tiwari2023,nitz2025,peng2025} and DINGO \cite{dax2021,dax2023}, which replaces Bayesian inference with neural posterior estimates for faster parameter estimation, are examples of the first approach.

In parallel, recent years have seen numerous applications of machine learning in the field of gravitational wave data analysis and inference \cite{xia2021,gabbard2022,ma2024,cuoco2025}. These applications range from noise reduction \cite{Ormiston:2020ele} and glitch detection \cite{Reissel:2025ykl,oshino2025} in the detector data stream to low-latency online event search \cite{mcleod2022,marx2024}. 
Another application of machine learning that enables faster likelihood calculations is numerical relativity surrogate models that can interpolate waveforms using a small number of numerical relativity simulations. Several other machine learning based models with partial or full implementation exist. These models are trained on different kinds of waveform approximations and are valid in various regions of the parameter space \cite{liao2021,lee2021,khan2021,chua2021,schmidt2021,thomas2022,shi2024,grimbergen2024,sun2025a,garg2025a,freitas2025}.
Yet a single, production-oriented framework that delivers on-the-fly gravitational waveform approximations over broad parameter ranges, at GPU speed, remains elusive \cite{cuoco2025}.

Largely, the current machine learning gravitational waveform approximation efforts suffer from the following limitations:
\begin{itemize}[nosep]
	\item Most models do not progress beyond the proof-of-concept stage or only a part of the waveform is modelled \cite{liao2021,lee2021,garg2025a}.
    \item Only a few source parameters in the parameter space are explored \cite{tissino2023}.
    \item Higher order modes, spin-precession effects, tidal deformability, etc., are not considered \cite{thomas2022}.
    \item Modelled waveforms are constrained to only certain kinds of systems, e.g., quasi-circular binary black holes \cite{khan2021}.
    \item Highly accurate numerical relativity surrogate or reduced-order models are not fast enough or have not been developed as a full-use implementation yet, though there is an increasing interest in this area \cite{gadre2024}.
    \item A unified broadly-adopted fast, and accurate waveform generation framework spanning all parameter regimes is still lacking \cite{sun2025a,cuoco2024,cuoco2025, freitas2025}.
\end{itemize}

In this paper, we study a machine learning model for generating fast and accurate effective one-body time-domain gravitational waveform approximations.
Our architecture is based on the best model of \citet{liao2021}, who used auto-encoder models for generating the inspiral and merger parts of the waveform from zero-spin binary black hole systems. We also use the auto-encoder approach, but consider full inspiral-merger-ringdown waveforms in aligned-spin systems. Our model is able to achieve reasonable accuracy (median of around $10^{-2}$) with orders-of-magnitude better speed than native waveform implementations.
Although the accuracy of this current model does not yet enable production-use, our results demonstrate that auto-encoder style machine learning frameworks can be used for generating full IMR gravitational waveform approximations at high speed. These kinds of fast but limited accuracy waveform models can be employed in cases where the likelihoods of a large number of waveforms need to be evaluated, for instance, during the initial parameter sampling stages to obtain approximate posteriors.
Furthermore, this paper is meant to be the first step towards the goal of a production-ready machine learning implementation of fast and accurate generation of gravitational waveform approximations.

The layout of the paper is as follows. \Sec{sec:data} outlines the data processing steps, while \Sec{sec:methods} describes our model architecture and training strategies. Then, in Section \Sec{sec:results}, we discuss our results. Finally, in \Sec{sec:summary} we conclude by outlining future directions and further applications of our model.

%%%%%%%%%%%%%%%%%%%%%%%%%%%%%%%%%%%%%%%%%%%%%%%%%%%%%%%%%%%%%%%%%%%%%%
%%%%%%%%%%%%%%%%%%%%%%%%%%%%%%%%%%%%%%%%%%%%%%%%%%%%%%%%%%%%%%%%%%%%%%

\section{Data Processing}
\label{sec:data}

We focus on aligned-spin binary black hole systems. Thus, our parameter space consists of four parameters: the two binary masses [$m_1$, $m_2$] and the $z$-components of their respective spins [$\chi_1(z),\chi_2(z)$]. We uniformly sample the black hole masses in the range $[5,75]\,\Msun$, with a hard mass ratio limit of $q = m_1/m_2 < 10$. For the spin parameters, we choose $\chi_1(z)$ and $\chi_2(z)$ as sampled uniformly in the prior range $[-0.99,0.99]$.

For each set of parameters, we calculate the full inspiral-merger-ringdown (IMR) waveform using the effective one-body approximant SEOBNRv4 from \texttt{lalsimulation} \cite{ligoscientificcollaboration2018} via \texttt{PyCBC} \cite{alexnitz2024} overlays. 
Although somewhat outdated in comparison with the newer SEOBNRv5 \cite{pompili2023}, this waveform approximant belongs to a family of aligned-spin time-domain waveforms calibrated with numerical relativity simulations for spin-precessing binary black holes systems. We only consider the leading order $h_{22}$ mode of the radiated gravitational wave. Each of our waveforms are \qty{1}{\second} long and are sampled at a rate of \qty{8196}{\hertz}. These are our input training data and waveform reconstruction targets.

%--- 
\subsection{Waveform Decomposition}

Now, the most important step in modelling any machine learning problem is proper data preparation and preprocessing. This becomes slightly tricky when we aim to train a model to generate highly accurate approximations to gravitational waveforms. It turns out that for most machine learning approaches, a full IMR time series is a relatively tougher problem to learn. The issue is exacerbated if we want to accurately extract features inherent in the original signal from the model generated waveform reconstructions.

\begin{figure}[t]
	\centering
	\includegraphics[width=\linewidth]{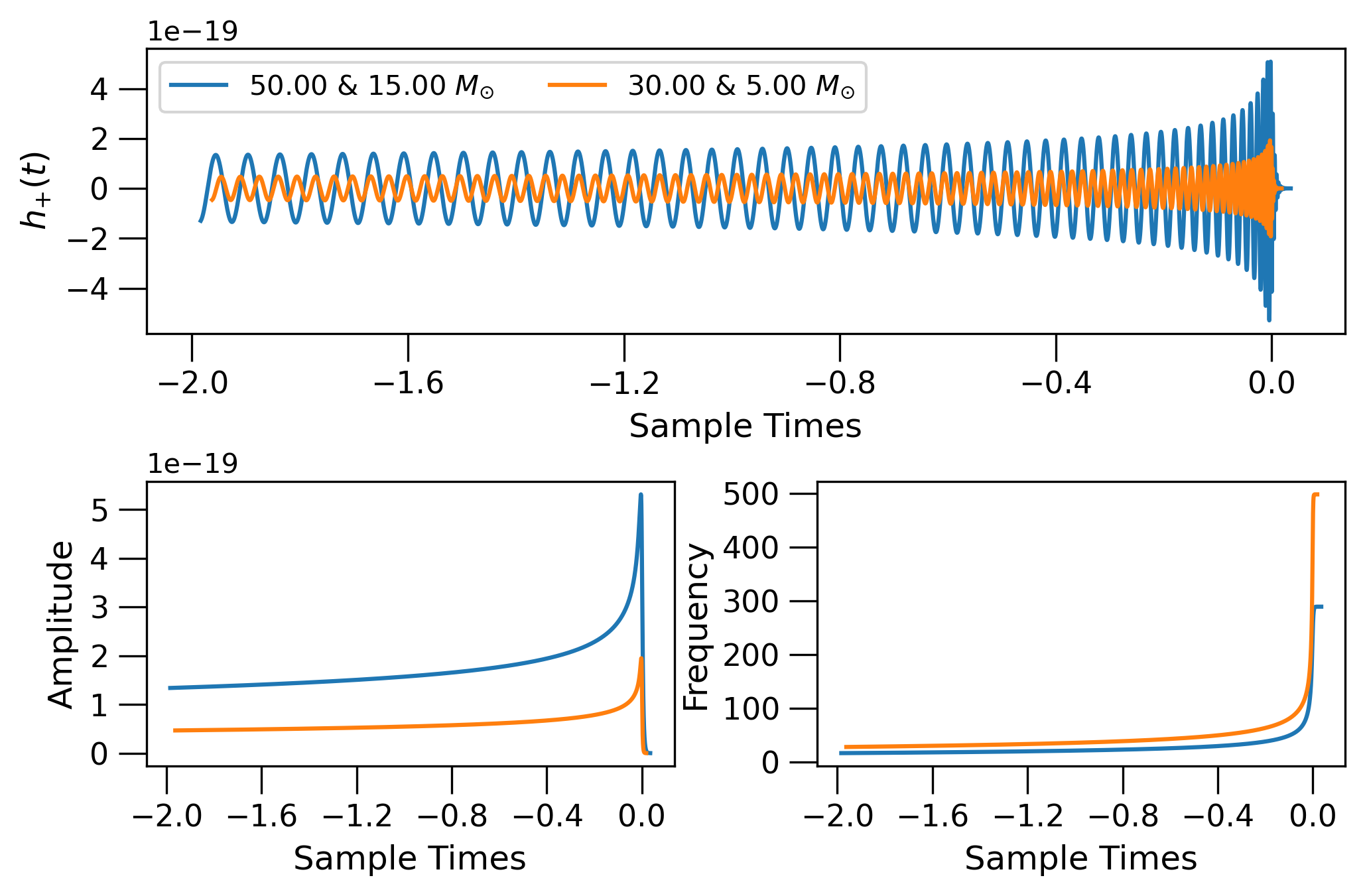}
	\caption{Representative plot of a gravitational waveform time series used for training and then generating targets later. A gravitational polarization waveform can be converted to an amplitude and frequency series.}
	\label{fig:data}
\end{figure}

Therefore, to make the target easier for our model to learn,  we note that
the leading order $h_{22}$ mode of the gravitational radiation can be decomposed as $h(t)=A(t) \ee^{-\aye\phi(t)}$, where $A(t)$ is the amplitude and $\phi(t)$ is the instantaneous phase of the waveform.
These are much simpler functions of time, and, for the aligned-spin case, exhibit strictly smooth, non-oscillatory behaviour. Thus, if we convert the IMR $h_{+,\times}(t)$ polarization time series into amplitude and instantaneous frequency series, the output targets become easier for our model to learn.
From polarizations $h_{+,\times}(t)$, the amplitude $A(t)$ and instantaneous frequency $f(t)$ can be obtained via the instantaneous phase $\phi(t)$ as,
\begin{align}
A(t) &= \sqrt{h_+^2(t) + h_\times^2(t)}\;, \nonumber \\
\phi(t) &= \text{unwrap}\,(\, \arctan \left(h_+(t),\,h_\times(t) \right)\; ), \nonumber \\
f(t) &= \frac{\phi(t+\Delta t) - \phi(t)}{2\pi\,\Delta t}\;,
\label{eq:ampfreq}
\end{align}
where $\Delta t$ is the time difference between two consecutive sampled data points.
Note that, by definition, this means the frequency series will have one  sampled data point less than the original polarization time series. \Fig{fig:data} shows a representative plot for the composition of a random input waveform data in our dataset.

We also normalize the amplitude and frequency series, so that the supplied input time series data to the model is within the same range of amplitudes and frequencies for all the waveforms:
\begin{align}
\widehat{A}(t) &= \frac{A(t) - \mu_A}{\sigma_A + \varepsilon}, \quad
\widehat{f}(t) = \frac{f(t) - \mu_f}{\sigma_f + \varepsilon}
\label{eq:keys}
\end{align}
where, $\widehat{A}$ and $\widehat{f}$ are the normalized quantities, $(\mu_A,\sigma_A)$ and $(\mu_f,\sigma_f)$ are the normalization factors and $\varepsilon$ is a small random number to keep values bounded near zero.
This normalization keeps the training scope bounded within certain limits for all input data, ensuring that our model produces reasonable outputs.

During the evaluation stage, after our model has learned, we will choose some desired input parameters and obtain the amplitude and frequency series as an output from the model. 
We will keep track of the normalization factors to denormalize the model outputs.
From these generated amplitude and frequency series, we will then calculate the phase and  reconstruct the polarizations, as follows:
\begin{align}
h_{+}(t) &= A(t) \cos \phi(t), \qquad
h_{\times}(t) = A(t) \sin \phi(t)\;.
\label{eq:polarizations}
\end{align}

%---
\subsection{Dynamic low-frequency cut-off}

Another issue with the input data is that sources with different binary component masses result in waveforms of varying durations. In principle, higher mass ratio binaries lead to shorter duration waveforms because the binary objects merge faster. However, our neural network model can only take fixed-length inputs. Thus, we need to ensure that our model receives only fixed-length inputs.

\begin{figure}[t]
	\centering
	\includegraphics[width=\linewidth]{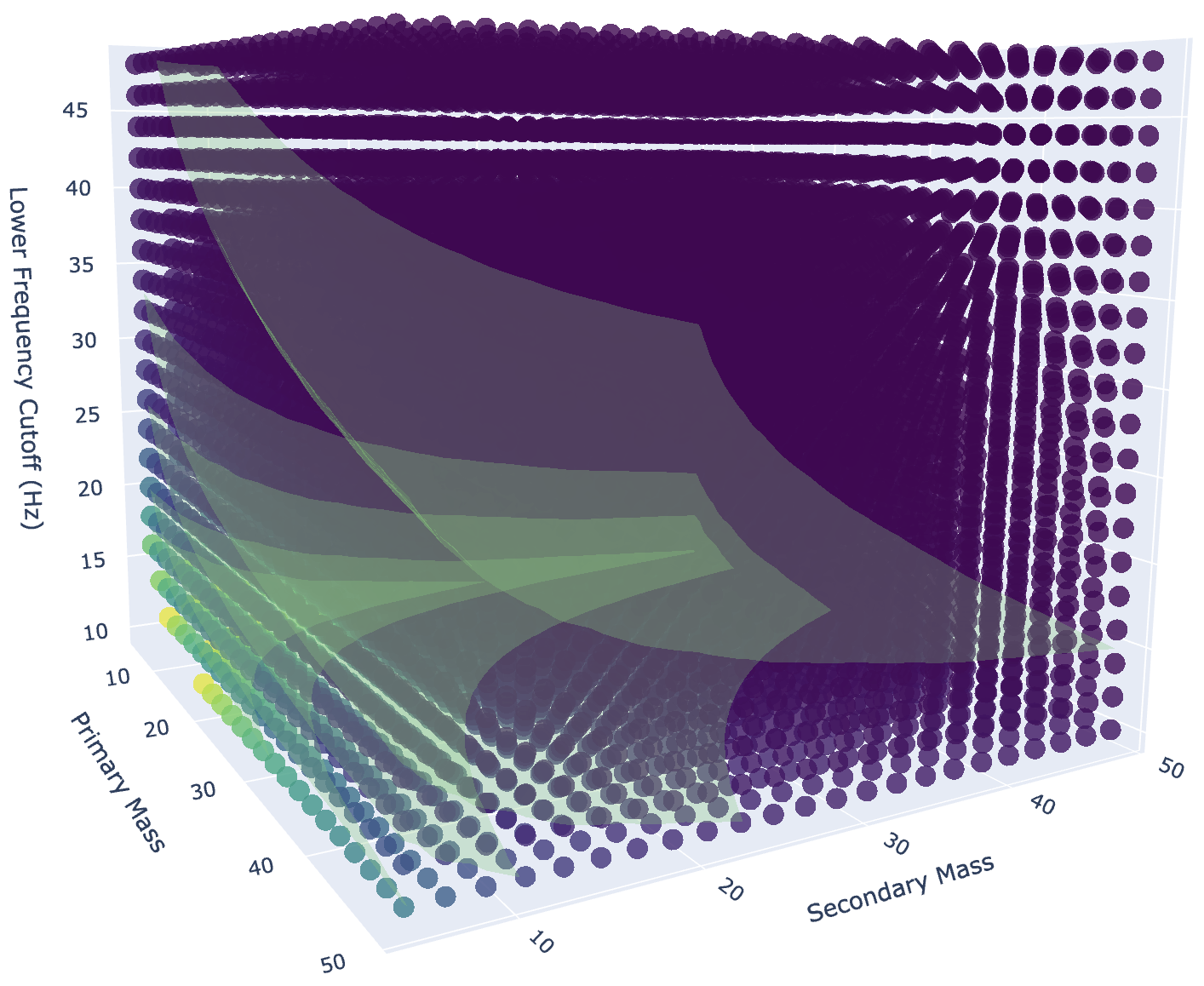}
	\caption{The duration of a waveform as a function of the binary component masses and the lower frequency cut-off, showing iso-surfaces of equal durations at 1s, 5s, 10s, 20s and 30s.}
	\label{fig:flow3d}
\end{figure}

A typical approach to ensure uniform length inputs is to append or prepend zeros at either end of the input data such that all inputs are of the desired length \cite{nam2019,islam2021a}. However, 
since this leads to non-physical discontinuities in the data, instead of prepending zeros to the input waveforms, we choose to have different lower frequency cut-off values for different input waveforms.
This ensures that even for high mass ratio binaries, we have sufficiently long polarization waveforms, since a smaller value of the cut-off frequency can be chosen.

To find the appropriate cut-off frequency, such that we have atleast the desired length of the waveform, we note that the time to coalescence of a binary black hole system under the quadrupole approximation is given by,
\begin{equation}
    t_c 
    \propto 
    f_{\rm low}^{-8/3} \mathcal{M}^{-5/3}\;,
\end{equation} 
where $f_{\rm low}$ is the lower frequency cut-off and $\mathcal{M} = (m_1 m_2)^{3/5} / (m_1+m_2)^{1/5}$ is the chirp mass \cite{cutler1994,sathyaprakash2009}.
\Fig{fig:flow3d} graphically illustrates this relation.
We estimate the proportionality constant $k$ by fixing $f_{\rm low}$ at \qty{40}{\hertz} for 500 waveforms with random chirp masses and calculating the time to coalescence. We then use this value of $k$ to calculate the required lower frequency cut-off for any new waveform with an arbitrary chirp mass, thus ensuring that the waveform has our desired duration. To account for the ringdown part of the full IMR waveform, we further reduce the obtained $f_{\rm low}$ by \qty{20}{\percent}. If the waveform exceeds our desired duration of one second, we finally truncate the initial phase of the early inspiral.

%%%%%%%%%%%%%%%%%%%%%%%%%%%%%%%%%%%%%%%%%%%%%%%%%%%%%%%%%%%%%%%%%%%%%%
%%%%%%%%%%%%%%%%%%%%%%%%%%%%%%%%%%%%%%%%%%%%%%%%%%%%%%%%%%%%%%%%%%%%%%

\section{Methodologies}
\label{sec:methods}

\subsection{Generative Model and Network Architecture}
\label{sec:model}

\begin{table*}[ht]
\caption{Details of the network architecture used in our 2C2E1D model.}
\label{tab:model}
\begin{tblr}{
  colspec = {X[2,l] X[2,l] X[4,l] X[3.5,l] X[2,l]},
}
\hline
\textbf{Component} & \textbf{pre-FC layers} & \textbf{CNN layers} & \textbf{post-FC layers} & \textbf{Notes} \\
\hline
XEncoder 
& None
& [Conv1D\,(in=1,out=5,k=16,d=1) + MaxPool\,(k=4)], 
[Conv1D\,(in=5,out=15,k=16,d=1) + MaxPool\,(k=4)];
& [Linear(in,500)+ReLU], 
[Linear(500,500)+ReLU], 
[Linear(500,out)+ReLU];
& labels concatenated to CNN output
\\
KeyEncoder 
& None
& None
& [Linear(in,500)+ReLU], 
[Linear(500,500)+ReLU], 
[Linear(500,out)+ReLU];
& 
\\
XConditional \& KeyConditional 
& None
& None
& [Linear(in,500)+ReLU], 
2$\times$[Linear(500,500)+ReLU], 
[Linear(500,out)+ReLU];
&
\\
Decoder 
& [Linear(in,800) + ReLU]
& [Conv1D\,(in=1,out=32,k=4,d=1) + MaxPool\,(k=4)], 
[Conv1D\,(in=32,out=64,k=4,d=2) + MaxPool\,(k=4)], 
[Conv1D\,(in=64,out=64,k=4,d=2) + MaxPool\,(k=4)];
& [Linear(in,800)+ReLU], 
[Linear(800,out)+ReLU];
& Inputs: $Z_X$(8), $Z_{\rm key}$(3), labels(4)
\\
\hline
\end{tblr}
\par\smallskip
Note: The values in the parentheses denote the size and other parameters of the respective module. `X' and `Key' components respectively handle the normalized input vector $[\widehat{A}(t),\widehat{f}(t)]$ and their normalization factors.
\end{table*}

\begin{figure}
    \centering
    \includegraphics[width=0.70\linewidth]{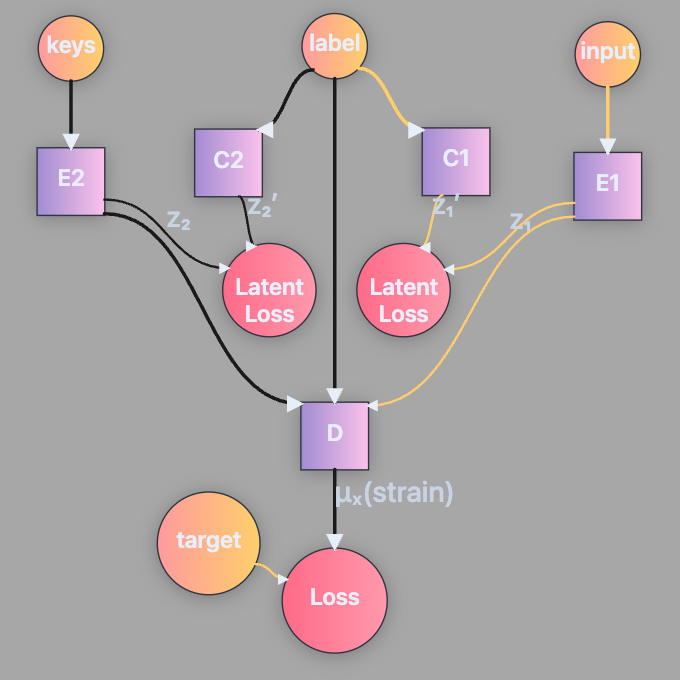}%
    \includegraphics[width=0.30\linewidth]{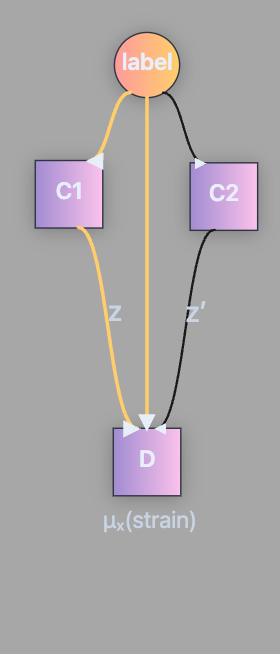}
    \caption{Architecture of the 2 conditionals, 2 encoder, 1 decoder (2C2E1D) auto-encoder model used in our analysis. The left side is the architecture used for training, while the right side is at the evaluation stage. Notice that the encoders are removed during evaluation and only the conditional parameter labels are used to generate outputs. See discussion for more details.}
    \label{fig:model}
\end{figure}

\begin{figure}
    \centering
    \includegraphics[width=\linewidth]{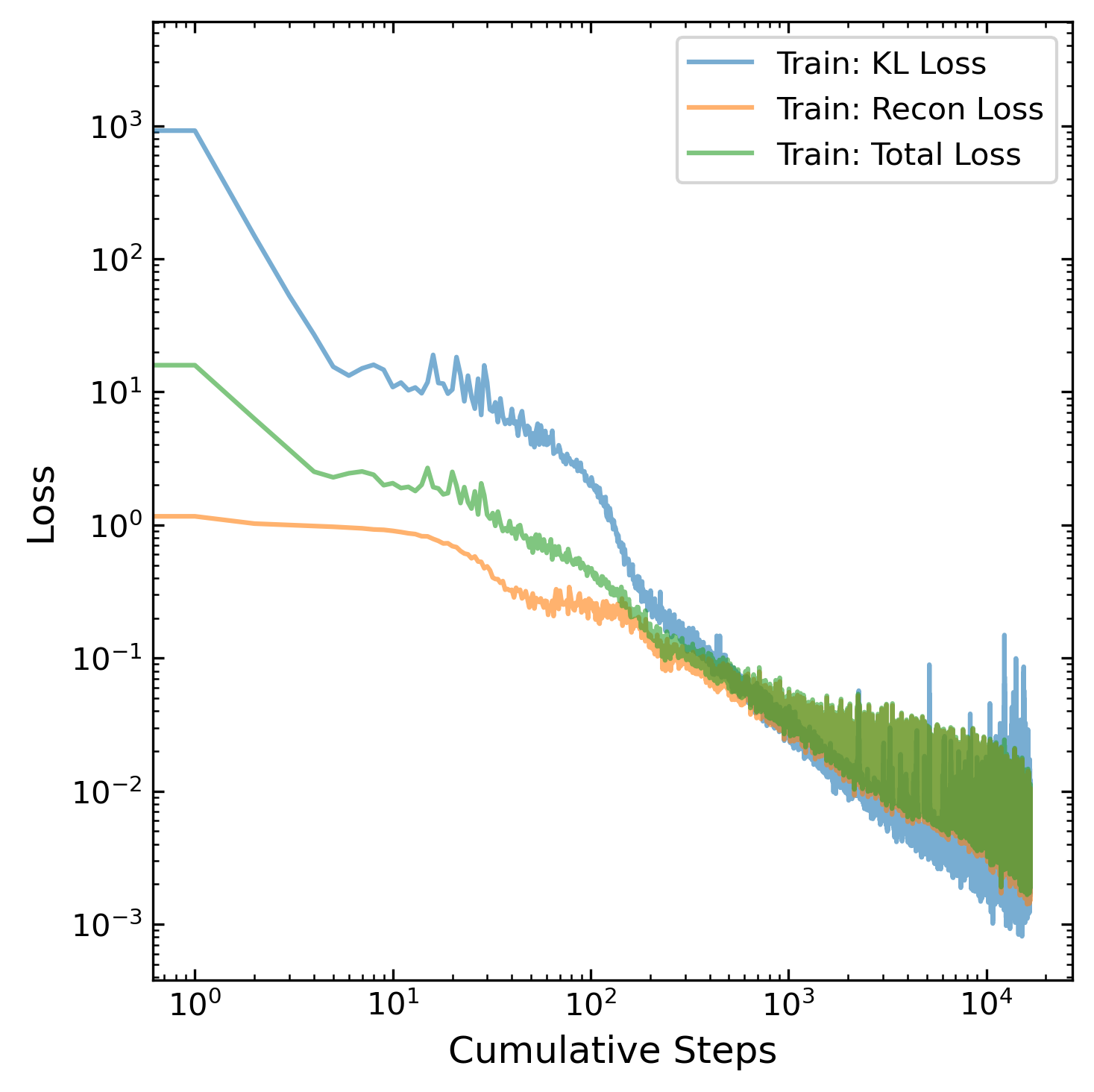}
    \includegraphics[width=\linewidth]{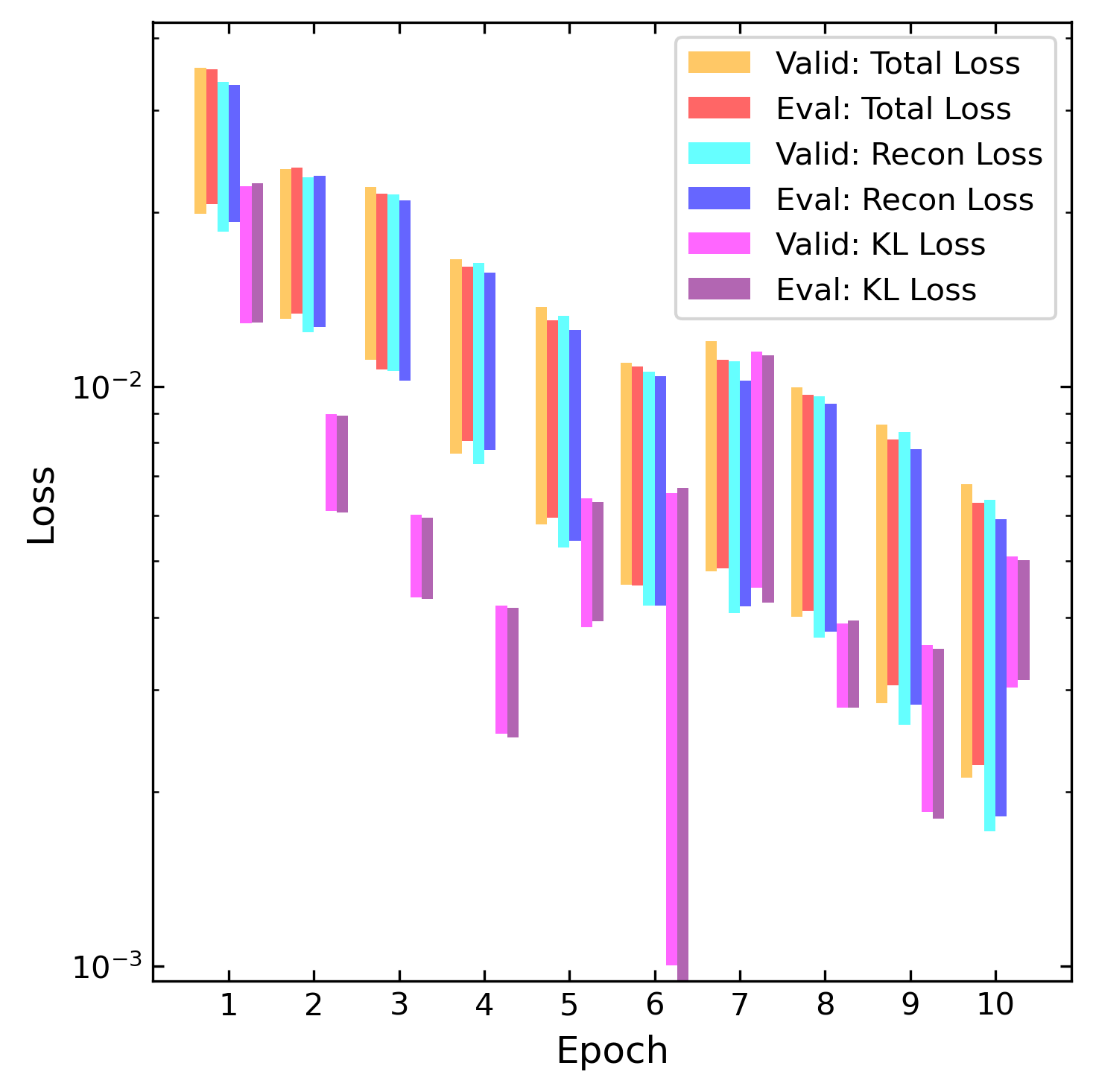}
    \caption{(top) Training loss as a function of number of training steps. One step is one batch of 50 waveforms passed to the model. This loss value is calculated while the model weights are being updated during the training process. (bottom)  Validation losses calculated on the validation set and the training set, at the end of each epoch, when the model training for that epoch is finished and the model weights are fixed. One epoch consists of 1669 training steps. The vertical bars have a height of two standard deviations and are centered at the mean value of the loss across different batches of data.}
    \label{fig:loss-time}
\end{figure}

Our task is to teach a machine learning model to learn and generate the gravitational waveform time series given some input source parameters. We find that an auto-encoder (AE)  model \cite{kingma2022} is the best-suited machine learning architecture for this purpose.
An AE model consists of an encoder and a decoder part, each of which can be an independent combination of a convolutional neural network and fully-connected linear layers. The encoder takes in some input data, encodes it into a latent space representation, and then the decoder attempts to decode this representation to produce a close semblance of the passed input data.

Furthermore, we aim for our model to generate waveforms that correspond to their specific set of parameters, given the binary source assumptions that we have made during the data preparation for model training. Therefore, we also need to supply in the source parameters as conditional priors to the model, alongside the respective input waveforms. The latent representation then encodes the information about the waveforms, while been conditioned on the input source parameters. After training the model, we will remove the waveform inputs but keep the conditional priors, so that the remaining model architecture can act as a generative model. Using this model we can now produce the required waveforms corresponding to any given input parameters. This type of an AE model is known as the conditional auto-encoder (CAE).

Now, given some parameters $\theta$ defining the intrinsic source configurations and the extrinsic detector specific properties, the corresponding waveform $h(t,\theta)$ can be determined completely. In other words, the forward problem $\theta\rightarrow h(t,\theta)$, from known parameters to the unknown waveform, is deterministic.
\footnote{Note that, degeneracies exist in the inverse problem of estimating parameters posteriors given some known waveform, in the sense that identical waveforms can belong to different sets of source parameters. An example for this is the case of waveforms from highly precessing binary systems that can have orbital timescale frequency modulations that are similar to those found in waveforms from systems with very low eccentricity non-circular orbits, thereby leading to degeneracy.
}
Therefore, if we choose a simple AE model, with uniform latent variable $z\sim\mu_{\mathrm{x}}$, for the latent space representation, our model is expected to work well. 
However, for our models to work effectively for waveforms from more complicated sources with many inherent features, we prefer a latent representation with the latent variable given by $z\sim\mu_{\mathrm{x}}+\sigma_{\mathrm{x}}\cdot\epsilon$, where $\epsilon\in\mathcal{N}(0,1)$ is a random number sampled from a normal distribution.
This also ensures that during the evaluation or inference stage, our model can interpolate in the parameter space, beyond the specific sets of parameters used for training. Such a model is now known as the variational auto-encoder (VAE), or with the conditional inputs included, as conditional variational auto-encoder (CVAE) \cite{kingma2019,doersch2021}.
Choosing to work with CVAE instead of CAE is a design choice at this current stage. We note that further studies will be required to test which design choice works best, although \citet{liao2021} found comparable performance for both their CAE and CVAE models. Lastly, the seed for the random number generator used to sample $\epsilon$ can be fixed at the start of model testing, so that identical test results are obtained at each run. For testing our trained model we fix the random number seed at 42. 

The specific architecture of our CVAE model consists of two conditional prior inputs, two encoders, and one decoder (2C2E1D), as illustrated in \Fig{fig:model}. The internal configuration of each of these modules is same as the 2C2E1D model in \citet{liao2021}, with one difference: both of our parameter label conditionals have 4 fully connected layers of size 500 each (see \Tab{tab:model} for more details).
We normalize each input waveform and pass the normalizing factors (mean and standard deviation) to the second encoder as keys. These keys are conditioned on their own conditional prior labels and help in denormalizing the reconstructed outputs during the inference stage.
More information about our variational model and the latent representation learning process is detailed in the Appendix.

%---
\subsection{Training Strategies}

Our full dataset consists of about $10^5$ waveforms and their source parameter data. We split this dataset into three mutually exclusive partitions for training, validation, and testing, with data fractions of \qty{70}{\percent}, \qty{10}{\percent}, and \qty{20}{\percent}, respectively. 
We then train our model for 10 epochs, using a batch size of 50, with random selections across the training set per epoch. The model weights are learned and optimized progressively during this training. At the end of each training epoch, we then fix the model weights (by setting the model to \texttt{eval} mode) and perform validation, first on the ``seen" data from training set and then on the ``unseen" data from validation set. \Fig{fig:loss-time} shows these loss functions during training and validation on different datasets. Our loss function is a sum of mean square error in the reconstructed amplitude and frequency series, and the KL divergence between the learned latent representations with \qty{10}{\percent} weight. More information on how the loss values are calculated at each step is outlined in the Appendix. Validation losses on both the training and validation sets are comparable and are lower than the training losses shown because validation is performed after the model has finished training for that epoch.

For both training and validation, we use a NVIDIA A100 \qty{80}{\giga\byte} GPU, with code written in \texttt{PyTorch} \cite{paszke2019} and run on a \texttt{CUDA} \cite{nickolls2008} environment.  Full training takes approximately \qty{30}{\hour} on this device. Our optimizer function is \texttt{ADAM} \cite{kingma2017} with an initial learning rate of $10^{-4}$ and a learning rate update schedule with $\gamma=0.1$ every 3 epochs, where $\gamma$ is the rate of decay.
We note that techniques like curriculum learning can be used to further enhance the training process \cite{garg2023a}.

\begin{figure*}[ht]
    \centering
    \includegraphics[width=\textwidth]{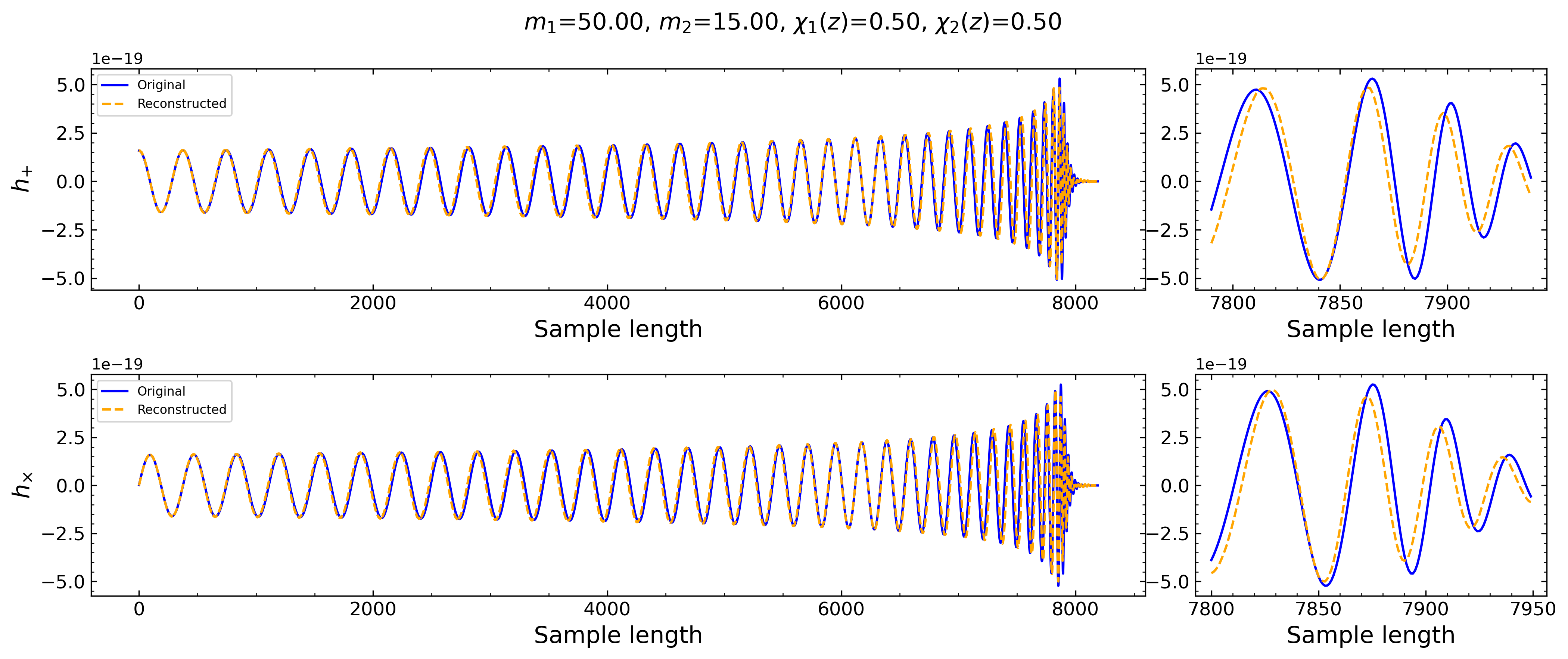}
    \caption{Overplot of the reconstructed waveform approximation with the original polarization time series for a randomly selected set of input parameters from the test dataset.}
    \label{fig:overplot}
\end{figure*}

%%%%%%%%%%%%%%%%%%%%%%%%%%%%%%%%%%%%%%%%%%%%%%%%%%%%%%%%%%%%%%%%%%%%%%
%%%%%%%%%%%%%%%%%%%%%%%%%%%%%%%%%%%%%%%%%%%%%%%%%%%%%%%%%%%%%%%%%%%%%%

\section{Results and Discussion}
\label{sec:results}

Once our model has been trained, we test the model performance using the test set, which consists of approximately \qty{20}{\percent} of 
the data in our full dataset. These waveforms for testing are chosen such that there are no repetitions from the training set, which ensures that the model does not overfit within a closed training scope. During the evaluation stage, we remove the encoders from the model and only use the conditionals that take in a set of input source parameters and pass them along to the decoder, which then generates our desired amplitude and frequency series. 
We denormalize these outputs using the normalization factors, and reconstruct the polarization time series using \Eq{eq:polarizations}.
A qualitative overplot of the original and reconstructed polarization time series is shown in \Fig{fig:overplot}.

%---
\subsection{Reconstruction Mismatch}

\begin{figure}
    \centering
    \includegraphics[width=\linewidth]{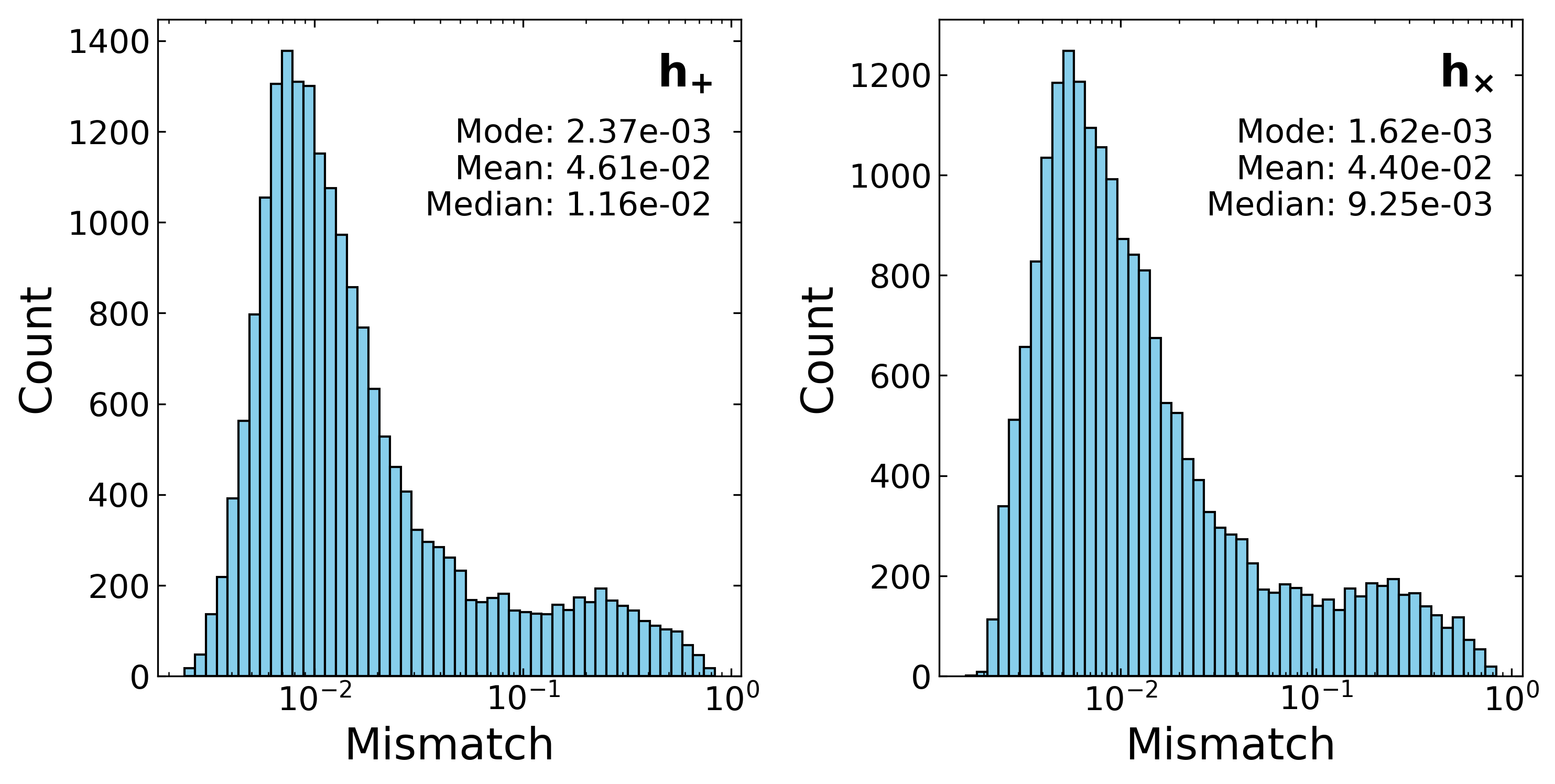}
    \includegraphics[width=\linewidth]{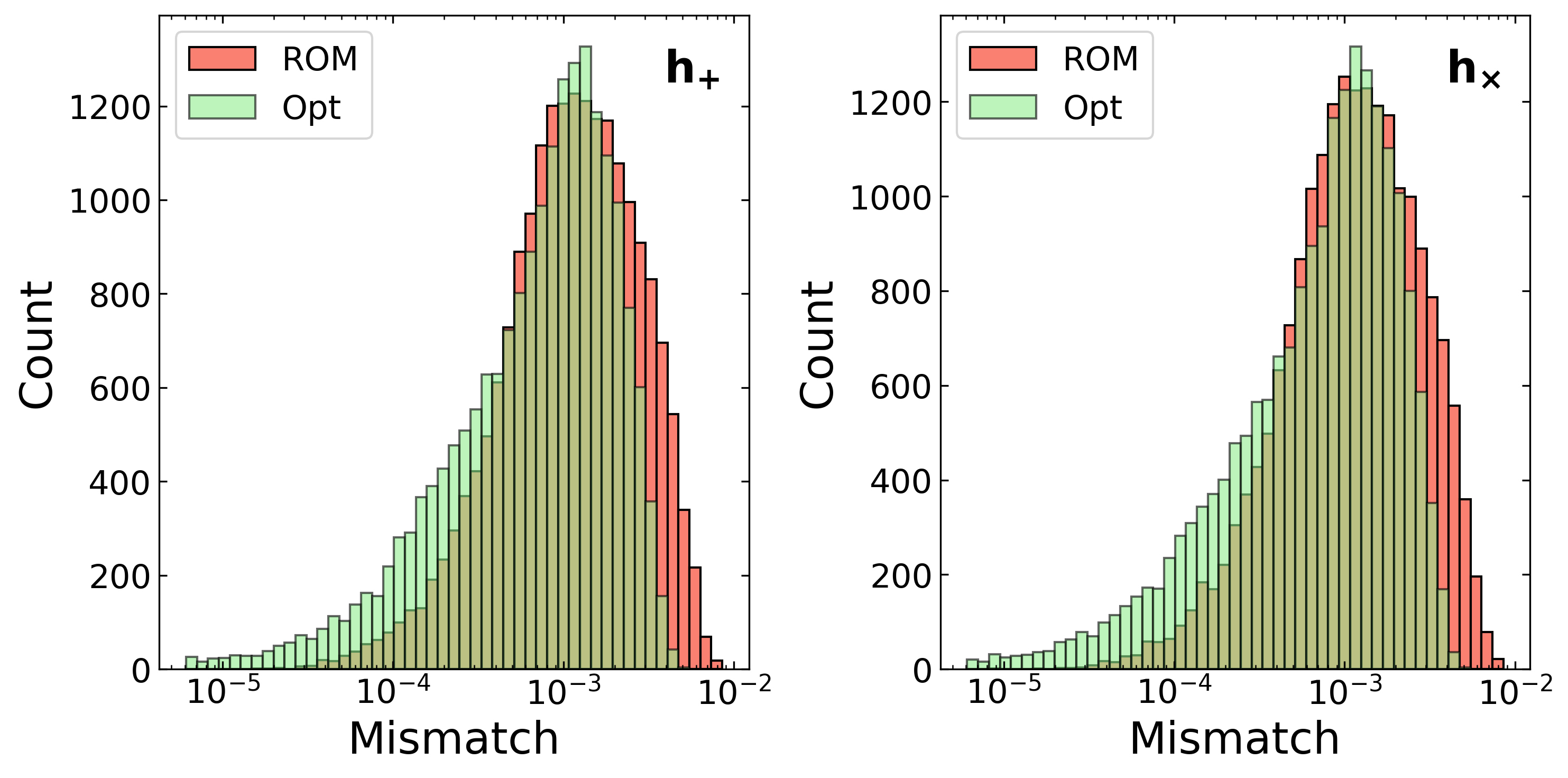}
    \caption{(top-panels) Histogram of the mismatch values for the reconstructed strain polarizations with the corresponding target waveforms in the test dataset. The parameter space spans the full training $\chi_{\rm eff}$ scope of [$-0.99,0.99$]. All waveforms have the same duration of the one second, while the lower frequency cut-off is dynamic. Mean, median, and mode statistics are also annotated. (bottom-panels) The mismatch value distribution for the ROM and opt variants of SEOBNRv4 waveform, for the same data in test dataset.}
    \label{fig:mmhist}
\end{figure}

To evaluate the accuracy of the waveforms generated by our model, we calculate the overlap and mismatch between the reconstructed waveforms and the original ones.
For the amplitude and frequency series, which are real-valued vectors, we can use the cosine distance as an evaluation criterion.
For any two discrete vectors $a$ and $b$, this is given by,
\begin{equation}
    \mathtt{D}(a,b) 
    = 1 - \dfrac{\langle a | b \rangle}{\sqrt{\langle a | a \rangle \langle b | b \rangle}}\;,
    \label{eq:mm-ampfreq}
\end{equation}
where $\langle a | b \rangle = \sum_i a_i b_i $,  is the inner product of the vectors and the quantity been subtracted on the right-hand side is the cosine similarity.

For the case of the polarization time series, we calculate the overlap between the original and reconstructed waveform by taking a noise-weighted inner product. 
Our reconstructed polarization time series, $h_{+}(t)$ and $h_{\times}(t)$, are real-valued functions of time. We do not combine these two components into a single complex-valued gravitational wave strain, and instead calculate the mismatch for each component separately.
For any two waveform $h_1(t)$ and $h_2(t)$, the noise-weighted inner product is given by,
\begin{equation}
    \langle h_1 | h_2 \rangle 
    = 4 \, \mathfrak{Re} \int_0^{\infty} \dfrac{\tilde{h}_1(f) \, \tilde{h}_2^{*}(f)}{S_n(f)} \diff f,
\end{equation}
where $\tilde{h}_1(f)$ denotes the Fourier transform of $h(t)$ and $\tilde{h}_2^{*}(f)$ the complex conjugate of $\tilde{h}_2(f)$. $S_n(f)$ is the detector noise, which we take to be given by the \texttt{aLIGOZeroDetHighPower} function from \texttt{lalsimulation} \cite{ligoscientificcollaboration2018}.
The optimal overlap (faithfulness) and the corresponding polarization mismatch of a reconstructed waveform $\hat{h}(t)$ with the original waveform $h(t)$ can then be calculated as,
\begin{equation}
    \tilde{\mathtt{MM}}(\hat{h},h)
    = 1 - \max_{\phi_c,\,t_c} \left[ \dfrac{\langle \hat{h} | h \rangle}{\sqrt{\langle \hat{h} | \hat{h} \rangle \langle h | h \rangle}} \right]\;.
    \label{eq:mm-hphc}
\end{equation}%

\begin{figure}
	\centering
	\includegraphics[width=0.5\linewidth]{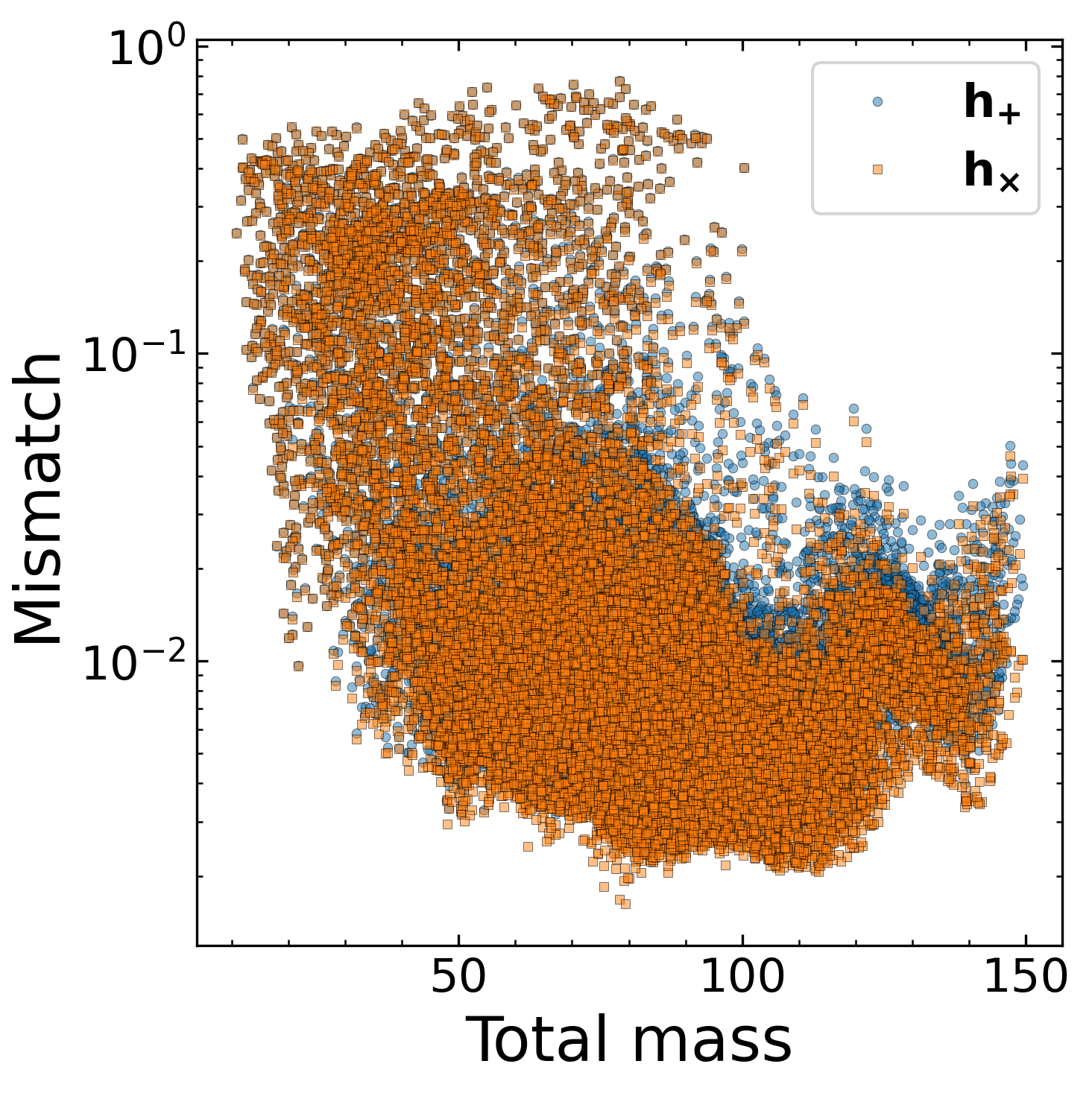}%
	\includegraphics[width=0.5\linewidth]{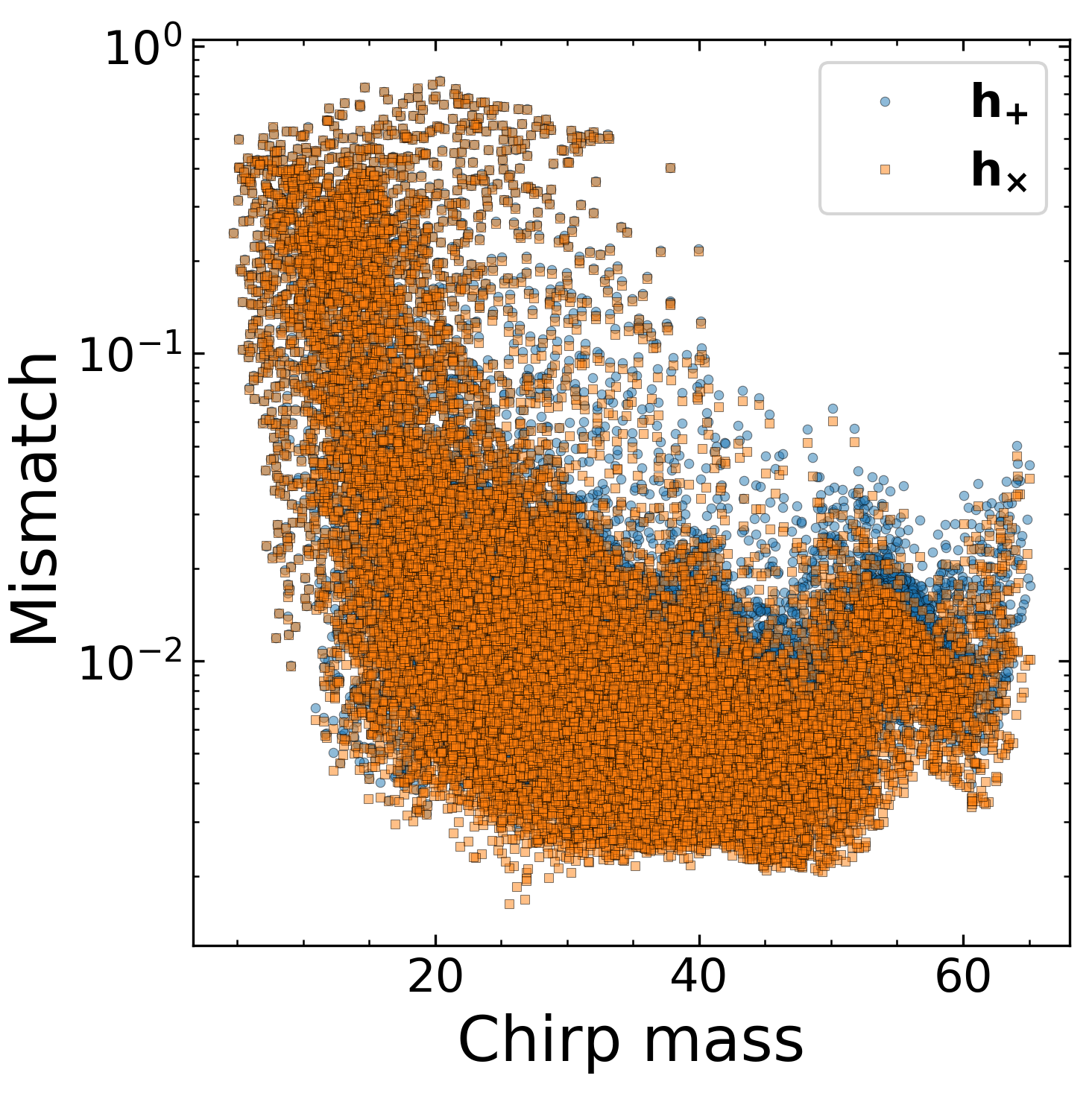}
	\includegraphics[width=0.5\linewidth]{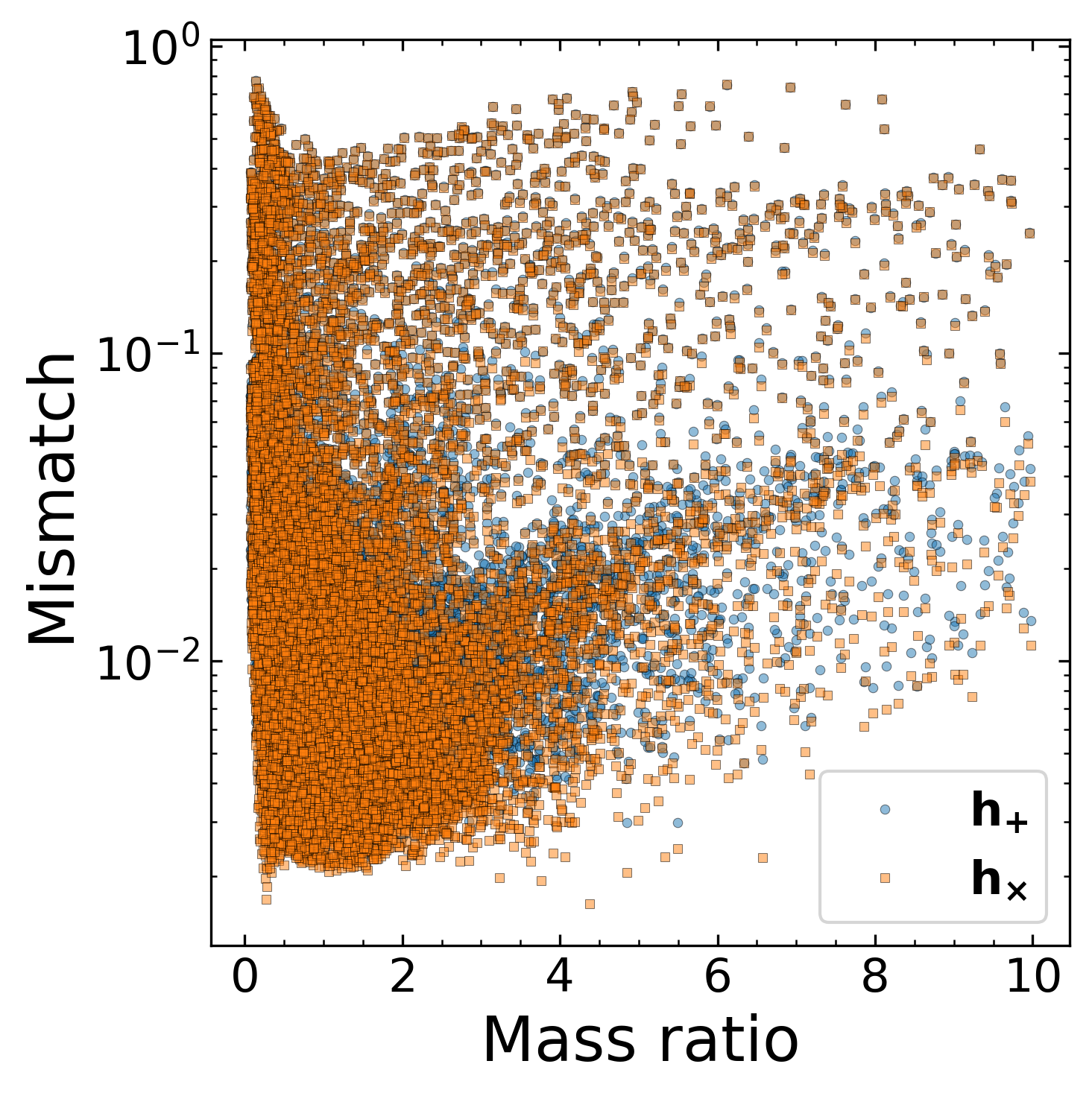}%
	\includegraphics[width=0.5\linewidth]{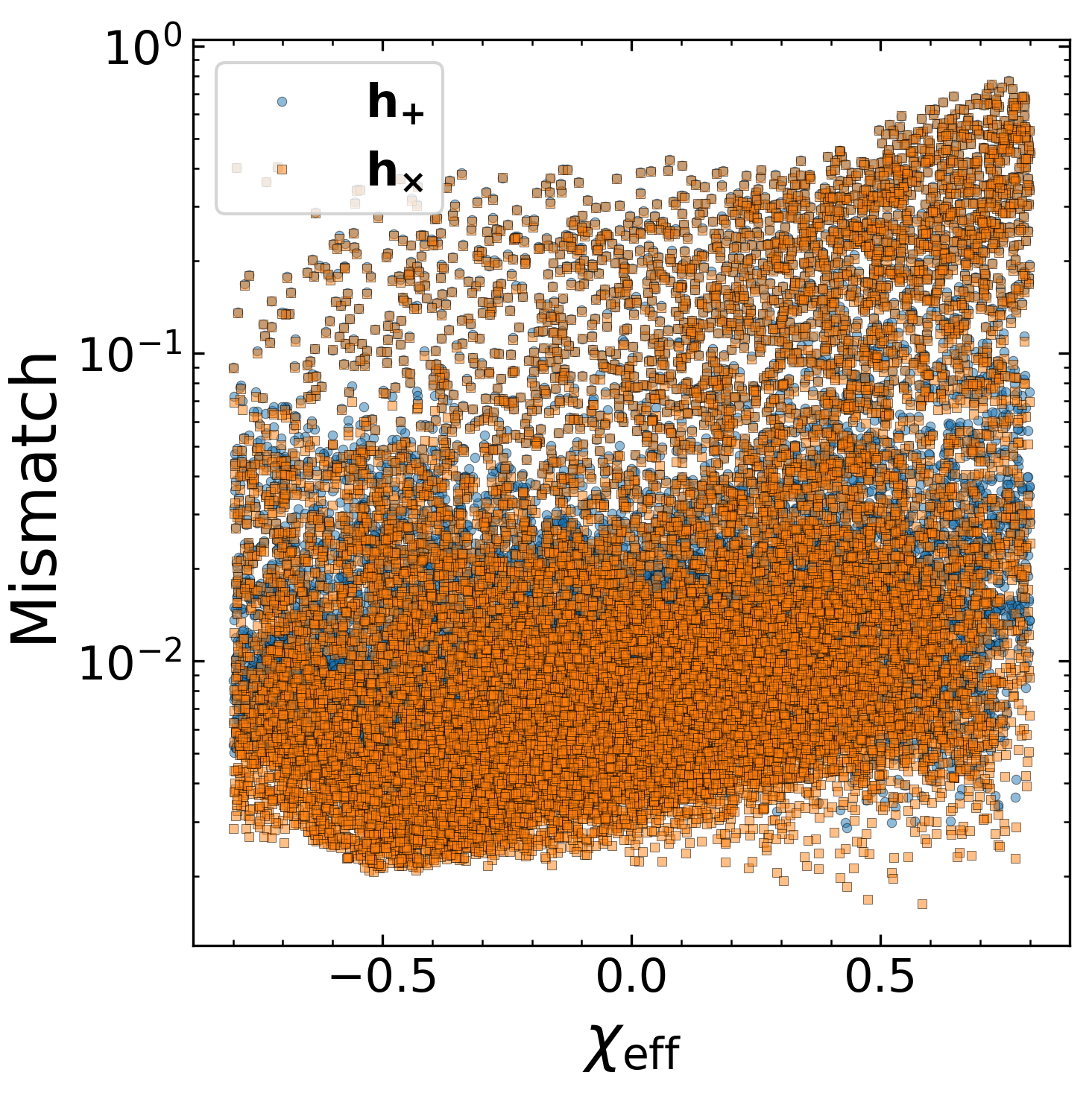}
	\caption{Mismatch values as a function of total mass (top left), chirp mass (top right), mass ratio (bottom left), and effective spin (bottom right) for reconstructed polarizations for all data in the test dataset. The parameter space is restricted to $\chi_{\rm eff}\in[-0.80,0.80]$.}
	\label{fig:mmplot}
\end{figure}

A summary of our results, namely the mean, median, best, and worst values of the mismatch, between the model generated waveforms and the originals, for all the data in the test dataset, is tabulated in \Tab{tab:results}. The polarization reconstructions are a derived quantity from the model generated amplitude and frequency series, and therefore, the accuracies of the latter are generally an order of magnitude better. \Fig{fig:mmhist} plots the histogram of the mismatch distribution across all the waveforms in our test set. It can be seen that the histogram peaks slightly to the left of $10^{-2}$ mark. 
However, several waveforms have extremely poor mismatch values.
The accuracy of ROM \cite{bohe2016} and opt \cite{devine2016,knowles2018} SEOBNRv4 waveform variants, obtained for the same test dataset at a fixed low-frequency cut-off of \qty{20}{\hertz}, is also shown in  \Fig{fig:mmhist} for comparison. They are non-machine-learning accelerated waveform versions of SEOBNRv4. The ROM variant, in particular, utilizes frequency-domain singular value decomposition to obtain reduced order basis representations of the gravitational waveform to make waveform generation faster \cite{purrer2014,purrer2016}.

We find that the worst mismatches from our model are obtained for waveforms that lie in the region of parameter space where the effective spin $\chi_{\rm eff}$ is large and positive. 
Recall that $\chi_{\rm eff}$ is the mass-weighted projection of the individual binary spins along the direction of the orbital angular momentum \cite{collaboration2025}. For the aligned-spin case, it is given by,
\begin{equation}
	\chi_{\rm eff} = \dfrac{m_1 \chi_{1z} + m_2 \chi_{2z}}{\chi_1 + \chi_2}\;.
\end{equation}%

\begin{table}[t]
	\caption{Summary of our results. 
		\label{tab:results}}
	\setlength{\tabcolsep}{4pt}
	\begin{tabular}{l*{4}{c}}
		\hline
		& \multicolumn{4}{c}{\textbf{Mismatch / cosine-distance}}  \\
		\cline{2-5}
		% \textbf{Quantity} 
		& \textbf{Mean} & \textbf{Median} & \textbf{Best} & \textbf{Worst} \\
		\hline
		$A(t)$   & $1.44\times10^{-3}$ & $5.99\times10^{-4}$ & $1.45\times10^{-4}$ & $ 7.82\times10^{-2}$  \\
		$f(t)$   & $1.15\times10^{-3}$ & $3.92\times10^{-4}$ & $1.06\times10^{-4}$ & $1.09\times10^{-1}$  \\
		$h_{+}$   & $4.61\times10^{-2}$ & $1.16\times10^{-2}$ 
		& $2.37\times10^{-3}$ & $8.32\times10^{-1}$ \\
		$h_{\times}$ & $4.40\times10^{-2}$ & $9.25\times10^{-3}$ 
		& $1.62\times10^{-3}$ & $8.34\times10^{-1}$ \\
		\hline
	\end{tabular}
	\par\smallskip
	Notes: Values for amplitude and frequency are calculated with \Eq{eq:mm-ampfreq}, while the polarization mismatches are computed using \Eq{eq:mm-hphc}.
\end{table}

\begin{figure}[t]
    \centering
    \includegraphics[width=\linewidth]{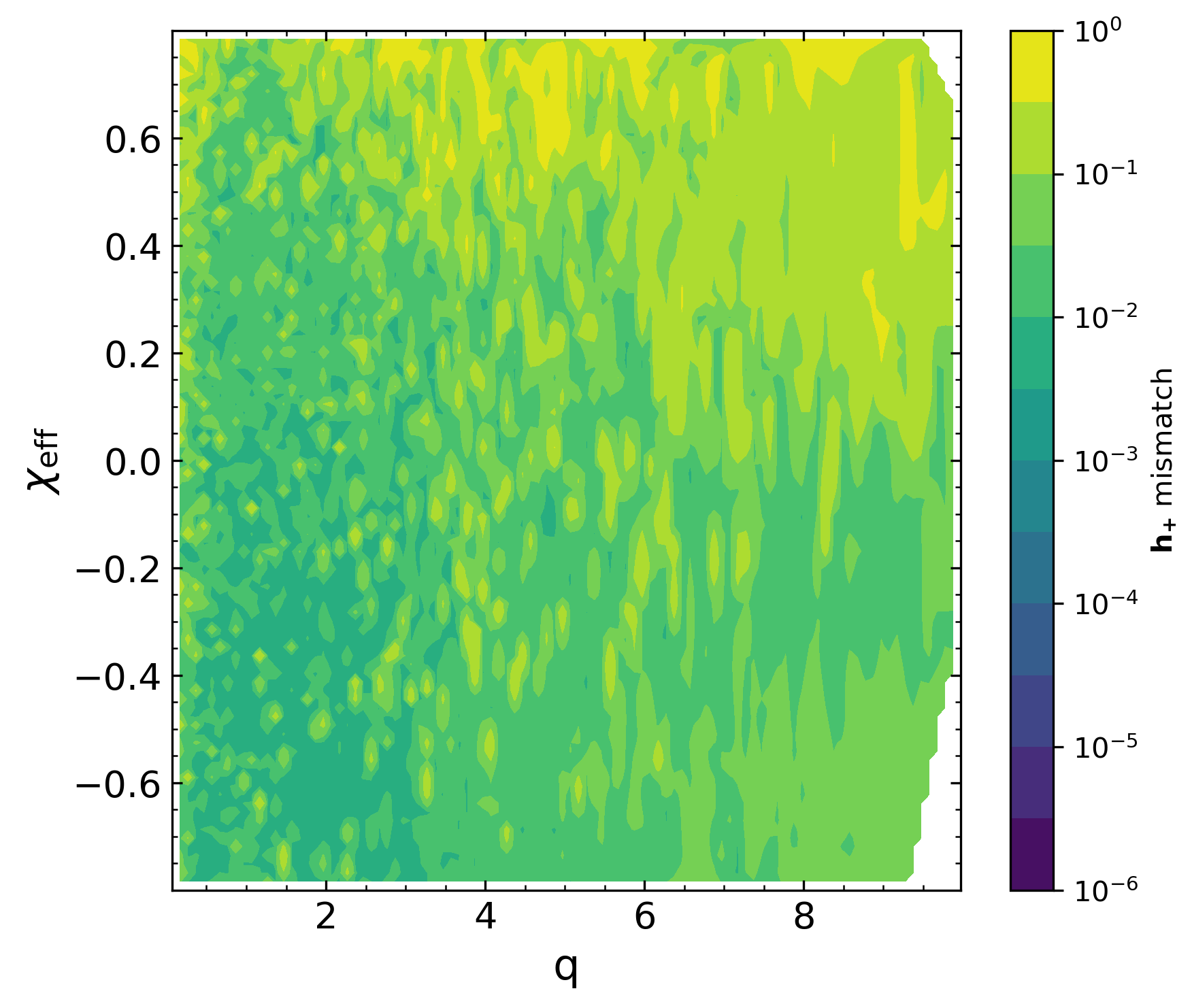}
    \caption{Contour plot of the mismatch values for $h_{+}$ reconstruction in the restricted mass ratio versus effective spin space of $\chi_{\rm eff}\in[-0.80,0.80]$.}
    \label{fig:contour}
\end{figure}

We also see a general trend of the mismatch values worsening as the positive $\chi_{\rm eff}$ or the mass ratio increases. With these outliers of our current model in mind, which could be fixed in a more refined version, we will restrict the parameter space to say, $\chi_{\rm eff}\in[-0.80,0.80]$. Thus, it is advisable to have this restricted parameter space when adopting our model for data analysis. \Figs{fig:mmplot}{fig:contour} plot the mismatch as a function of mass and effective spin for this restricted space. These plots show the tomography over the parameter space for the accuracy of the waveforms generated by our model.

Although restricted, the parameter space currently supported by our model, should be extensive enough to cover most gravitational wave events. Observationally, LVK population studies till date have found that the effective inspiral spin is typically small in magnitude and is centered near zero, with values of $\chi_{\rm eff}\gtrsim0.5$ uncommon in the present catalog \cite{collaboration2025}. This supports our choice of a restricted parameter space in the production run of our model and it's actual adoption.

We want to emphasize that the poor mismatch in some of the generated waveforms can be attributed to errors in accurately reconstructing the peak in the amplitude (and phase) series and the subsequent ringdown portion of the waveform by the model. Namely, the peak amplitude value is often not reached by the reconstructed waveforms, leading to poor mismatch cases. An effective way to remedy this could be to have a model work in two parts: one that focuses solely on the inspiral portion and the other for the merger and ringdown. This will be one possible avenue we could explore in a follow-up work.

%---
\subsection{Waveform Generation Efficiency}

\begin{figure}[t]
    \centering
    \includegraphics[width=\linewidth]{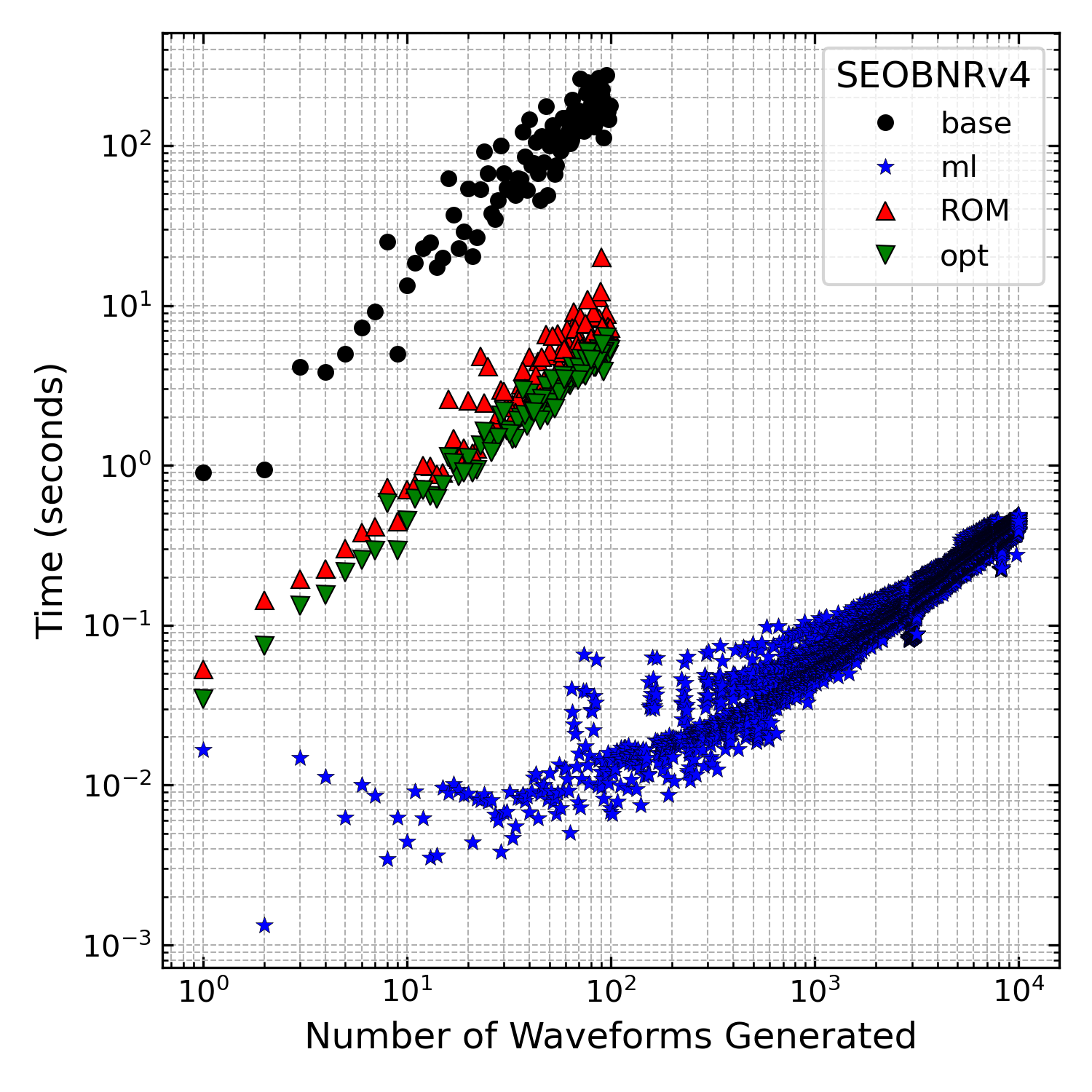}
    \caption{Time taken to generate different number of waveforms with random parameters using our trained model on a GPU. The waveform generation time using the base, ROM and opt variants of SEOBNRv4 implementations are also plotted for comparison, with $f_{\rm low}$ fixed at 20 Hz for all waveforms.}
    \label{fig:speed}
\end{figure}

We verify the speed of waveform generation for our trained model by performing iterations of $n$ number of waveforms generated for random inputs parameters, where $n$ ranges from $1$ to $10^4$. We use the same Nvidia A100 \qty{80}{\giga\byte} GPU that we trained our model on for generating these waveforms.
 graphically illustrates the comparison of waveform generation speed using our model and the native implementation in \texttt{lalsimulation} available via \texttt{PyCBC} \cite{alexnitz2024} and \texttt{SWIG} \cite{wette2020} wrappers.
It can be easily seen from \Fig{fig:speed} that our model is 
able to generate $10^3$ waveforms in $\sim10^{-1}$ seconds. On average, this generation speed is about 200--300 times faster than the native waveform implementations (the Python wrappers for \texttt{lalsimulation} waveforms have negligible overheads). 
The mean of per-waveform generation time across all the iterations is \qty{50}{\micro\second}. As the number of waveforms generated increases, the time taken by the native implementation is proportional to the number of waveforms. In contrast, the waveform generation time using the machine learning model remains almost constant till 100 waveforms generated, until when the model reaches full GPU usage capacity. 
At the start, the machine learning models require some warm-up time during which sufficient memory is identified on the GPU and \texttt{cuda} instruction sets are loaded. In order to enable fair comparison, in \Fig{fig:speed} we perform a model warm-up prior to actual use by generating 10 batches of 2 dummy waveforms which are discarded.  
It is only after more than 100 waveforms are generated at a time that we see the machine learning model exhibit compute times proportional to the number of waveforms.
In this regime, we see 4 orders of magnitude speed-up over the native waveform implementation. Even with the faster ROM and opt variants of SEOBNRv4, our machine learning model achieves 2--3 orders of magnitude better speed. The performance is expected to further enhance when using a better-class GPU in future.

Another advantage of a machine learning model like ours for generating waveform approximations is that during the inference stage, all computations can be performed on the GPU. Once a pre-trained waveform generation model is deployed for online use, it can be loaded directly onto the GPU and generate output waveforms for any input parameters on the same device. This eliminates the need for CPU-to-GPU communication and provides an additional speed-up for online parameter inference tasks. 
However, it should be noted that that the GPU usage is purely optional and the same trained model can be loaded onto a CPU and used for waveform generation on that device. Even on a CPU, the efficiency of a machine learning waveform generation model is expected to be better, since the model only performs elementary matrix multiplication operations to obtain the outputs.
The key advantage of a GPU accelerated machine learning model is the batch-wise bulk waveform generation. This is especially distinctive at large number of waveforms generated, for instance, more than 100 waveforms, as seen from \Fig{fig:speed}. Now, traditional MCMC sampling codes work sequentially and call for waveforms one at a time. However, the \texttt{emcee} sampler package does support vectorized lists of parameter inputs and computed outputs \cite{foreman-mackey2013}. This could be easily made to work with the batches of waveforms generated from our model.

%---
\subsection{Latent Space Uncertainty}

\begin{figure}[t]
    \centering
    \includegraphics[width=\linewidth]{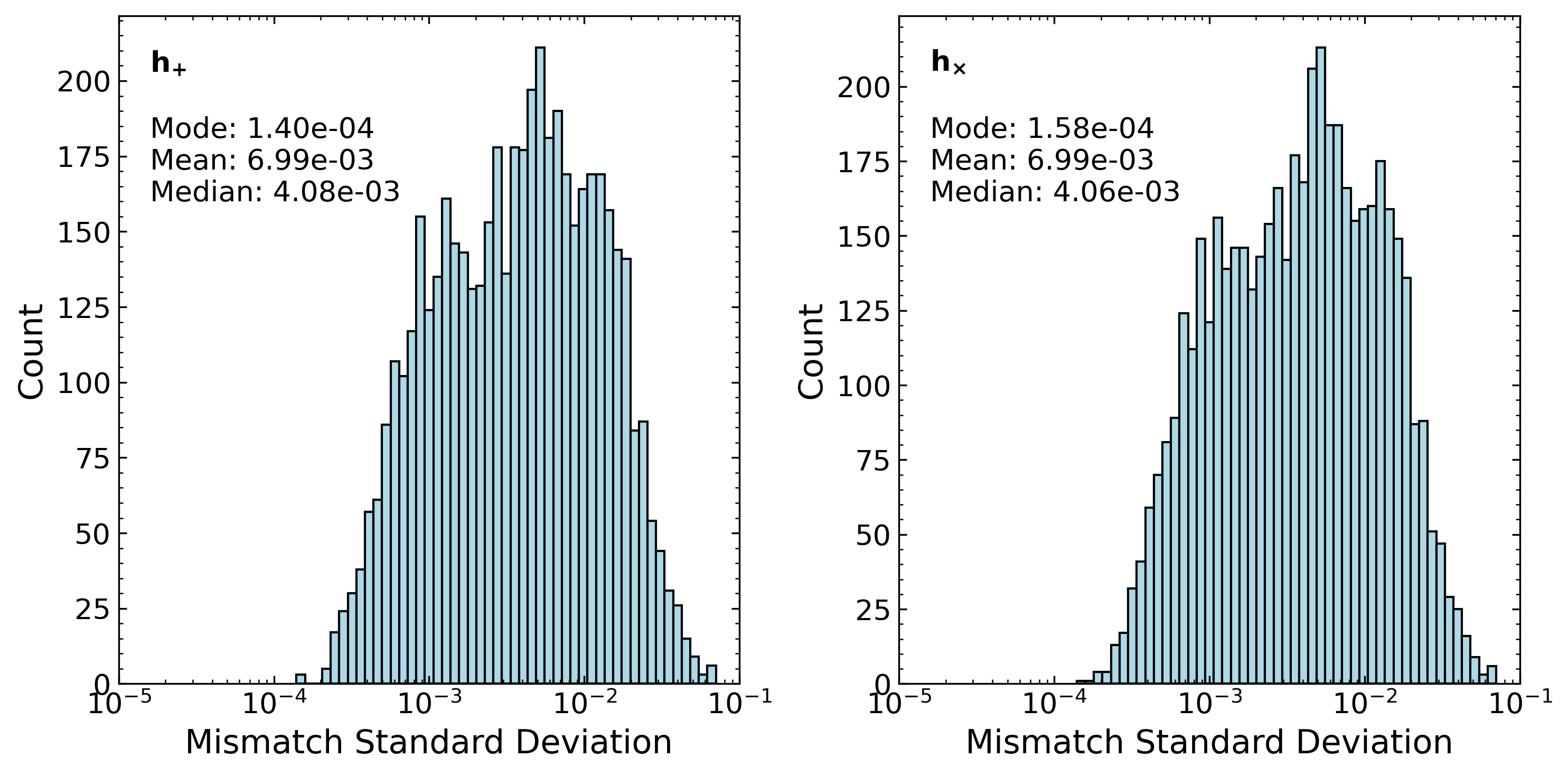}
    \caption{Distribution of the mismatch standard deviation value for 100 generations of a  waveform iterated over 5000 random waveforms selected from the test dataset.}
    \label{fig:uq}
\end{figure}

As we know, the waveform generation problem is inherently deterministic. So, our trained model should generate the exact same waveform reconstruction for a set of input parameters, every time the model is used.
However, for reasons described in \Sec{sec:model}, we chose our latent representation to follow a random distribution, so that our generative model can interpolate within the parameter space outside the specific sets of parameter used for training. This random distribution in the latent space adds uncertainty to the waveform generated by our variational model, when variables are sampled from the latent space in the decoder step.

We can quantify this latent sampling uncertainty for the model by supplying the same parameters $n$ times and obtaining $n$ waveform approximations using the trained model. We then calculate the mean and the standard deviation in the mismatch value of these $n$ waveforms in comparison to the actual SEOBNRv4 target waveform for the same parameters. If the generated waveform is always the same, as would be the case for an encoder which learns to associate each input to a fixed-length latent vector instead of a distribution, the mismatch values would be identical each time. However, in a variational model, these $n$ mismatch values are expected to have a spread. Further, the mismatch of the model generated waveform varies a great deal across the parameter space. Thus, we need to calculate the standard deviation in the mismatch for $N$ waveforms with randomly selected parameters from the parameter space.
\Fig{fig:uq} plots the histogram of the distribution of standard deviation in the mismatch for $n=100$ generations of a waveform iterated for $N=5000$ random waveforms. We find that the latent sampling uncertainty of our model can thus be constrained at a median mismatch standard deviation value of $4\times10^{-3}$.

%%%%%%%%%%%%%%%%%%%%%%%%%%%%%%%%%%%%%%%%%%%%%%%%%%%%%%%%%%%%%%%%%%%%%%
%%%%%%%%%%%%%%%%%%%%%%%%%%%%%%%%%%%%%%%%%%%%%%%%%%%%%%%%%%%%%%%%%%%%%%

\section{Summary and Conclusion}
\label{sec:summary}

In this paper, we have presented a conditional variational auto-encoder model, based on the best performing architecture of \cite{liao2021}, for faster generation of aligned-spin effective one-body SEOBNRv4 gravitational waveform approximations.
For training the model, we take the gravitational $h_{+,\times}(t)$ polarization waveforms for some sets of [$m_1$,$m_2$,$\chi_1(z)$,$\chi_2(z)$] parameters and convert them to amplitude and instantaneous frequency series. We then train and optimize our model to accurately generate the normalized amplitude and frequency series, given the input source parameters and normalization factors, by minimizing the reconstruction loss. To use the trained model for waveform generation, during the evaluation or inference stage, we remove a part of the model architecture, pass the source parameters as inputs, and obtain the generated amplitude and frequency series. Then, using the normalization factors, that we kept a track of, we reconstruct the desired polarizations as outputs. 

Our training parameter space spans a mass range of $[5,75]\,\Msun$ and spin values in $[-0.99,0.99]$. Our trained model is able to generate $10^3$ waveforms in $\sim10^{-1}$ seconds with a median polarization mismatch of about $10^{-2}$ for the reconstructed waveforms. The latent sampling uncertainty of our variational model can be quantified at a median mismatch standard deviation of $4\times10^{-3}$. Most of the poor mismatch reconstructed waveforms correspond to parameters at the boundary of the parameter space, and in the restricted space of $\chi_{\rm eff}\in[-0.80,0.80]$ our model has better accuracy.

In summary, our current work aims to be an instructive step towards full development of production-ready machine learning models for fast and accurate generation of gravitational waveform approximations. Our model is orders-of-magnitude more efficient that both the native implementation of the SEOBNRv4 waveform and its ROM and opt variants. However, we note that the accuracy of the model presented here is not sufficient enough to allow a drop-in replacement in likelihood calculation and parameter estimation processes. 
In spite of the limited accuracy of generated output waveforms, our model should be usable for specific cases where high-volume of theoretical waveform approximations are required and where speed and parameter range coverage are more crucial than the exact accuracy of each theoretical waveform. For instance, one such use-case could be rapid estimation of an approximate credible sky localization region for gravitational merger events \cite{singer2016}.

There are numerous ways to increase the accuracy of the machine learning model further, e.g. by fine-tuning network design, or by increasing the number of training waveforms. In particular, since our present work is meant to be preliminary model to demonstrate that encoder-decoder networks can satisfactorily approximate a full IMR waveform, we employed a simple mean squared error loss to measure the reconstruction error. Optimization of the loss function, so that it includes the mismatch information, could be another approach to increase the output accuracy of the generated waveforms. However, again in this case the model architecture will have to be optimized and just a change in the loss function does not improve the model performance. We can also modify the output targets of the model to unnormalized amplitude and frequency series, so that the dependency on the normalization factors is removed. 
Lastly, numerous further extensions and applications of our encoder-decoder style machine learning framework for waveform generation are in sight. These include expanding the parameter space to include precessing and eccentric orbit parameters, a two-part model focusing on inspiral and merger-ringdown waveforms separately, etc.

%%%%%%%%%%%%%%%%%%%%%%%%%%%%%%%%%%%%%%%%%%%%%%%%%%%%%%%%%%%%%%%%%%%%%%
%%%%%%%%%%%%%%%%%%%%%%%%%%%%%%%%%%%%%%%%%%%%%%%%%%%%%%%%%%%%%%%%%%%%%%

\label{acknowledgements}
\begin{acknowledgments}
We would like to thank Deep Chatterjee (MIT Kavli) for insightful discussions and remarks during his visit to Japan.
S.G. is grateful to the gravitational waves research groups at Osaka Metropolitan University and Seoul National University for fruitful discussions. 
Meeting with Alistair McLeod (UWA) was helpful in interpreting the results. Meetings with HM Lee, Nicolas Chartier, Arif Shaikh, Elaha Khalouei at SNU, Chunglee Kim at Ewha and Alvin Li (UTokyo) were helpful in framing the methodologies for taking our work forward.
Training of our model was performed on the LDG-CIT cluster at Caltech and RESCEU-BBC cluster at UTokyo. The work of FLL is supported by Taiwan's National Science and Technology Council (NSTC) through Grant No.~112-2112-M-003-006-MY3.
\end{acknowledgments}

% \nocite{*}
\bibliographystyle{apsrev4-2}
\bibliography{mlpaper}  % Produces the bibliography via BibTeX.

%%%%%%%%%%%%%%%%%%%%%%%%%%%%%%%%%%%%%%%%%%%%%%%%%%%%%%%%%%%%%%%%%%%%%%
%%%%%%%%%%%%%%%%%%%%%%%%%%%%%%%%%%%%%%%%%%%%%%%%%%%%%%%%%%%%%%%%%%%%%%

%\newpage
%\clearpage

\appendix*
\section{2C2E1D CVAE model 
\label{sec:appendix}}

We use a 2-conditional, 2-encoder, 1-decoder model for generating gravitational waveform approximations corresponding to a specific set of parameters. Our task can be formally stated as follows.
Given some inputs $\mathbf{x} = [A(t),f(t)]_{t\in\{1...T\}}$ consisting of normalized amplitude and instantaneous frequency series, some intrinsic source parameter labels $\mathbf{y}=[m_1,m_2,\chi_{1}(z),\chi_2(z)]$, and the normalization factors (called keys) $\mathbf{s} = [\mu_A,\sigma_A,\mu_f,\sigma_f]$, we want to find the reconstructed outputs $\widehat{\mathbf{x}}$, such that the difference between $\mathbf{x}$ and $\widehat{\mathbf{x}}$ is minimized.
Note that in a future production run, the reconstruction targets will be the unnormalized amplitude and frequency series, but in our present model, the target is $\mathbf{x}$.

The set of inputs/keys and labels is passed to the two encoders, which produce the posterior distributions
$\mathbf{Z}_1 \sim q_{\phi_1}(\mathbf{z}\!\mid\!\mathbf{x},\mathbf{y})$ and
$\mathbf{Z}_2 \sim q_{\phi_2}(\mathbf{z}\!\mid\!\mathbf{s},\mathbf{y})$,
while the two conditional priors yield reference samples
$\mathbf{Z}'_1 \sim p_{\psi_1}(\mathbf{z}\!\mid\!\mathbf{y})$ and
$\mathbf{Z}'_2 \sim p_{\psi_2}(\mathbf{z}\!\mid\!\mathbf{y})$.
These distributions are,
\begin{align}
q_{\phi_1}(\mathbf{z} \mid \mathbf{x}, \mathbf{y}) &= \mathcal{N}\!\left(\mathbf{z};\, \mu^{(q)}_1(\mathbf{x},\mathbf{y}),\, \mathrm{diag}\big(\sigma^{(q)}_1(\mathbf{x},\mathbf{y})^2\big)\right)\;, \nonumber \\
q_{\phi_2}(\mathbf{z} \mid \mathbf{s}, \mathbf{y}) &= \mathcal{N}\!\left(\mathbf{z};\, \mu^{(q)}_2(\mathbf{s},\mathbf{y}),\, \mathrm{diag}\big(\sigma^{(q)}_2(\mathbf{s},\mathbf{y})^2\big)\right)\;, \nonumber \\
p_{\psi_j}(\mathbf{z} \mid \mathbf{y}) &= \mathcal{N}\!\left(\mathbf{z};\, \mu^{(p)}_j(\mathbf{y}),\, \mathrm{diag}\big(\sigma^{(p)}_j(\mathbf{y})^2\big)\right), \,\, j=1,2\;.
\end{align}%

During the decoding stage, 
we draw a sample from each of the two latent spaces, say $\mathbf{Z}_1$ and $\mathbf{Z}_2$, and concatenate them with the labels $\mathbf{y}$ to obtain $\mathbf{z}=\mathbf{Z}_1+{\mathbf{Z}_2}+\mathbf{y}$, which is passed to the decoder. 
We also alternatively used the mean of posteriors  $\mathbf{z}=\tfrac{1}{2}(\mathbf{Z}_1{+}\mathbf{Z}_2) + \mathbf{y}$, as the input to decoder, however the training performance remained the same.
The reparametrization trick
$\mathbf{z}=\boldsymbol{\mu}(\cdot)+\boldsymbol{\sigma}(\cdot)\odot\boldsymbol{\epsilon}$ with $\boldsymbol{\epsilon}\!\sim\!\mathcal{N}(\mathbf{0},\mathbf{I})$
is used so gradients can flow through stochastic sampling with low variance. The decoder then outputs the
reconstruction $\widehat{\mathbf{x}}=\mathbf{f}_{\mathrm{decoder}}(\mathbf{z},\mathbf{y})$.

The training process aims to minimize both the reconstruction error and the Kullback–Leibler (KL) regularization between the encoder posteriors and the conditional priors.
The decoder likelihood, that we obtain the correct input $\mathbf{x}$ as the output during the reconstruction using a latent sample $\mathbf{z}$ and a conditional 
label prior $\mathbf{y}$, is given by,
\begin{multline}
p_\theta(\mathbf{x} \mid \mathbf{z}, \mathbf{y}) 
\\
= \prod_{t=1}^{T} \mathcal{N}\!\left(\mathbf{x}_t;\, \mu^{(x)}_\theta(t \mid \mathbf{z}, \mathbf{y}),\, \mathrm{diag}\big(\sigma^{(x)}_\theta(t \mid \mathbf{z}, \mathbf{y})^2\big)\right).
\end{multline}%

The conditional log-evidence decomposes as,
\begin{equation}
\log p_\theta(\mathbf{x} \mid \mathbf{y})
\ge \mathbb{E}_{q_\phi}\!\left[\log p_\theta(\mathbf{x} \mid \mathbf{z},\mathbf{y})\right]
- \mathrm{KL}\!\left(q_\phi\,\|\,p_\psi\right), 
\end{equation}
where the right-hand side of the inequality is the evidence-based lower bound (ELBO). We want to maximize ELBO, which is equivalent to minimizing a sum of (i) the negative expected log-likelihood and (ii) the KL regularization for the latent space representations.

With our Gaussian decoder $p_{\boldsymbol{\theta}}(\mathbf{x}\mid \mathbf{z},\mathbf{y})$, the negative log-likelihood is (up to constants) a weighted MSE,
\begin{equation}
-\log p_{\boldsymbol{\theta}}(\mathbf{x}\mid \mathbf{z},\mathbf{y})
\propto \sum_{t,d}\frac{\big(x_{t,d}-\mu^{(x)}_{\boldsymbol{\theta},d}(t\mid \mathbf{z},\mathbf{y})\big)^2}{\sigma^{(x)2}_{\boldsymbol{\theta},d}(t\mid \mathbf{z},\mathbf{y})},
\end{equation}
which yields the following reconstruction loss when $\boldsymbol{\sigma}^{(x)}_{\boldsymbol{\theta}}$ is fixed (plain MSE):
% or learned (heteroscedastic weighting):
\begin{align}
\mathcal{L}_{\mathrm{recon}} &= \left\lVert \mathbf{x} - \widehat{\mathbf{x}}\right\rVert^2
= \left\lVert \mathbf{x} - f_{\mathrm{decoder}}(\mathbf{z},\mathbf{y})\right\rVert^2 \;.
\end{align}%

On the other hand, the KL regularization refers to the KL divergence between two probability distributions, defined as
\begin{equation}
\mathrm{KL}\,\!\big(q(\mathbf{z})\,\|\,p(\mathbf{z})\big)
= \mathbb{E}_{\mathbf{z}\sim q}\!\left[\log \frac{q(\mathbf{z})}{p(\mathbf{z})}\right]\;.
\label{eq:kl-def}
\end{equation}%

For the case of two univariate $d$-dimensional Gaussian distributions with $\mathbf{\sigma}_1,\mathbf{\sigma}_2>0$ (as in our model), the KL regularization can be reduced to,
\begin{multline}
\mathrm{KL}\,\!\Big(\mathcal{N}(\boldsymbol{\mu}_1,\operatorname{diag}\boldsymbol{\sigma}_1^{2})
\,\Big\|\, 
\mathcal{N}(\boldsymbol{\mu}_2,\operatorname{diag}\boldsymbol{\sigma}_2^{2})\Big)
\\
= \frac{1}{2}\sum_{i=1}^{d}\!\left[
\frac{\sigma_{1,i}^{2}}{\sigma_{2,i}^{2}}
+\frac{(\mu_{2,i}-\mu_{1,i})^{2}}{\sigma_{2,i}^{2}}
-1+\log\!\frac{\sigma_{2,i}^{2}}{\sigma_{1,i}^{2}}
\right].
\end{multline}
If one of the distributions is a normalized Gaussian $\mathcal{N}(\mathbf{0},\mathbf{I})$, this further becomes $\frac{1}{2}\sum_{k=1}^{d}\!\big(\mu_k^2+\sigma_k^2-1-\log\sigma_k^2\big)$.

Our overall objective function can now be compactly written as, 
\begin{equation}
\mathcal{J}(\theta,\phi_{1:2},\psi_{1:2})
= \mathcal{L}_{\mathrm{recon}}
+\beta\cdot\mathcal{L}_{\mathrm{KL}} \;,
\end{equation}%
where, the KL regularization consists of two encoder–encoder symmetric terms and four encoder–prior terms,
\begin{align}
\mathcal{L}_{\mathrm{KL}}
=& \tfrac{1}{2} \,\left[
\mathrm{KL}\,\!\big(q_{\phi_1}\|q_{\phi_2}\big)
+\mathrm{KL}\,\!\big(q_{\phi_2}\|q_{\phi_1}\big)
\right] \nonumber
\\
 & + \sum_{i=1}^{2}\sum_{j=1}^{2}
\mathrm{KL}\,\!\big(q_{\phi_i}\,\|\,p_{\psi_j}\big).
\end{align}%
We choose $\beta=0.1$ during our training process and try to optimize $(\theta,\phi_{i=1,2},\psi_{j=1, 2})$ so that our objective $\mathcal{J}$ is minimized. A relatively low weight for the KL divergence term reduces the latent sampling uncertainties of the model.

%%%%%%%%%%%%%%%%%%%%%%%%%%%%%%%%%%%%%%%%%%%%%%%%%%%%%%%%%%%%%%%%%%%%%%

\end{document}